\documentclass[iop]{emulateapj}
\usepackage{graphics}
\usepackage{amssymb,amsmath}
\usepackage{epsfig}
\begin{document}

\title{Hydrodynamic Models of Radio Galaxy Morphology: Winged and X-shaped Sources}

\author{Edmund~J.~Hodges-Kluck$^{1}$ \& Christopher~S.~Reynolds$^{1}$} 
\affil{Department of Astronomy, University of Maryland, College
Park, MD 20742-2421, USA}
\email{ehodges@astro.umd.edu}

\slugcomment{Accepted by ApJ 3/22/11}
\shorttitle{Hydrodynamic Models of Radio Galaxy Morphology}
\shortauthors{Hodges-Kluck \& Reynolds}

\begin{abstract}
We present three-dimensional hydrodynamic models of radio galaxies interacting
with initially relaxed hot atmospheres and explore the significant off-axis
radio lobe structures which result under certain conditions.  With a focus on
the ``winged'' and ``X-shaped'' radio galaxy population, we confirm the
importance of observed trends such as the connection of wing formation with
jets co-aligned with the major axis of the surrounding atmosphere.  These
wings are formed substantially by the deflection of lobe plasma flowing back
from the hot spots (backflow) and develop in two stages: 
supersonic expansion of an overpressured cocoon at early times followed by 
buoyant expansion at later times.  We explore a limited parameter space of 
jet and atmosphere properties and find that the most prominent wings are 
produced when a decaying jet is injected into a small, dense, highly elliptical 
atmosphere.  On the basis of this search, we argue that the deflection of
backflow by gradients in the hot atmosphere is a strong candidate for forming
observed wings but must work in tandem with some other mechanism for forming
the initial wing channels.  Our models indicate that lobe interaction
with the hot atmosphere may play a dominant role in shaping the morphology of
radio galaxies.
\end{abstract}


\section{Introduction}

\begin{center}
\begin{figure}[t]
\vspace{0.5cm}
\centerline{
\includegraphics[scale=0.4]{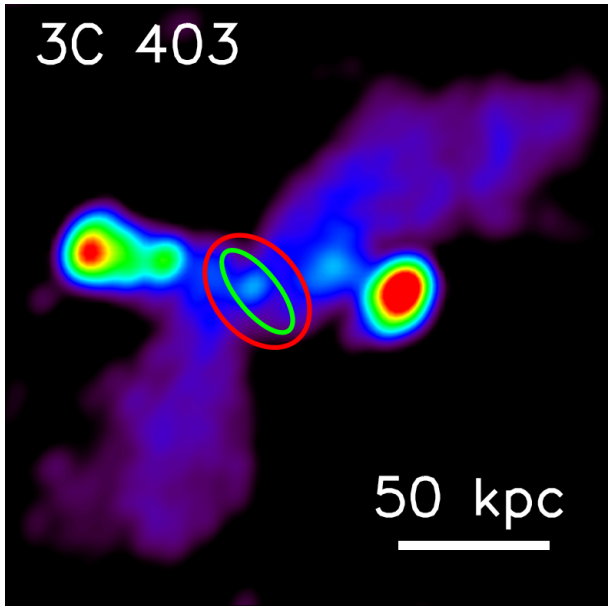}
}
\figcaption{{\small 3C~403, a typical XRG showing long, faint wings emanating from the
same core as the edge-brightened lobes (east--west axis).  The red and
green ellipses have the ellipticity and position angle of the host galaxy
and ISM \citep[from][]{kraft05} respectively (not to scale). }}
\label{3c403}
\end{figure}
\end{center}

Recently, X-shaped radio galaxies (XRGs)---a peculiar class of radio galaxies
with two pairs of misaligned lobes \citep{ekers78,leahy92}---have received significant 
attention as new observations have explored and critiqued the competing
models for the origin of their odd morphology.  The distinctive 
``X'' shape occurs due to the intersection of two centro-symmetric
synchrotron-emitting structures at a common nucleus (e.g.~Figure~\ref{3c403}). 
One of these structures is associated with an active relativistic jet (the ``primary''
lobes) whereas the other is fainter, more ragged, and does not
appear to harbor a jet (the ``secondary'' lobes or ``wings'').  The wings
can be long, collimated, and symmetric about the nucleus, and may even exhibit
Z-shaped morphology of their own \citep{gopal03}.  The origin of the wings is
not presently understood, but XRGs share other characteristics which indicate
a common origin.

X-shaped sources make up about 5--10\% of Fanaroff-Riley Type~II
\citep[FR~II;][]{fan74} radio galaxies \citep[e.g.][]{leahy92}.  The 
consensus sample is fairly small, but with the identification of $\sim 100$
``candidate'' XRGs \citep{cheung07}, a search for trends is now possible.  
Based on the combined consensus and candidate sample, XRGs tend to have
radio powers and host optical luminosities near the FR~I/II ``break'' in a 
Ledlow-Owen radio power--visual magnitude \citep{ledlow96} type plot
\citep{cheung09,landt10}, and reside in elliptical galaxies with larger-than-average
black hole masses \citep[inferred from the $M-\sigma$ relation and broad
lines from the active galactic nucleus;][]{mezcua10}.  XRGs do not seem to be
in galaxies currently undergoing a merger based on inferred starburst history
\citep{mezcua10} and a spectroscopic search for broad emission lines and
dusty nuclei \citep{landt10}, but the dynamic ages of their active lobes 
are younger than the age of the starburst \citep{mezcua10}.  The wings are
preferentially co-aligned with the \textit{minor} axis of the host, whereas the
primary lobes tend to be co-aligned with the major axis \citep{capetti02,
saripalli09}; this alignment has been confirmed in the X-rays which trace the
distribution of hot gas in and around the host galaxies that provides
the working surface which shapes the radio galaxy \citep{hodges10a}.
The radiative age of the wings does not seem to follow a clear trend, as some
wings have steeper spectral indices than the primary lobes (implying ageing)
whereas others do not \citep{lal05}.  Notably, some wings are actually longer
than the primary lobes, and wings are only found in radio galaxies with strong
bridges.

A critical review of the observational data and existing models is found in
\citet{gopal10b}; here we briefly summarize the main threads.  The origin of
the secondary lobes has been attributed to (i) a rapid reorientation of the
spin axis of the supermassive black hole (SMBH) powering the jet (i.e. the
wings are fossils), either due to a SMBH merger \citep[e.g.][]{rottmann01,zier01,merritt02}
or rapid precession \citep[e.g.][]{dennett02}; 
(ii) redirection and collimation of ``backflow'' (spent jet
plasma flowing back from the terminal shocks) \citep{leahy84,worrall95,
capetti02,kraft05}; (iii) a binary AGN \citep{lal07}; and 
(iv) interaction of the jet with disturbed morphology \citep[e.g. stellar shells, phase-wrapped remnants of
a merged disk galaxy;][]{gopal10b}.  Since XRGs are usually strongly bridged
sources and are apparently aware of their environments, it is worth examining
closely the hypothesis that the X-shaped morphology originates from an 
interaction between the radio galaxy and its environment.  
In this paper, we seek to test the viability of the \textit{backflow model} with 
three-dimensional hydrodynamic simulations in elliptical atmospheres.

In the backflow model, wings are produced by a single AGN outburst with
powerful jets as the backflow is diverted along the steepest pressure gradient
of the surrounding atmosphere (i.e. the minor axis).  The wings either
rise buoyantly or are driven in this direction \citep{leahy84,worrall95,kraft05} or
form as supersonic outflows along the direction of least resistance from a 
ruptured overpressured cocoon \citep{capetti02,zanni03}.  These scenarios
naturally explain the observed correlation between the wings and the minor axis as
well as the dearth of strictly FR~I XRGs (the weaker FR~I sources do not 
produce powerful back-flows).  However, the backflow model faces significant challenges, most
notably the very long wings in many XRGs (often longer than the primary
lobes).  Whereas powerful jets are expected to drive through the surrounding
medium supersonically, wings which expand buoyantly would do so at most
transonically for most of their lives \citep{leahy92}.  

Because, in this model, fluid effects are primarily responsible for the 
``X'' shape, hydrodynamic simulations are an ideal proving ground.  
Although recent hydrodynamic simulations of radio galaxy morphology exist in 
the literature, \citep[e.g.][]{sutherland07,falceta10,gaibler10}, most are not
concerned with the formation of lateral asymmetries such as wings.  In view of
advances in simulating radio galaxies in the past decade, the mounting evidence 
that XRGs constitute a population demands a critical look at the backflow model
with new simulations.  In this
paper, we discuss 3D hydrodynamic models of radio galaxies ignited in initially
relaxed, elliptical atmospheres and examine how wings form. 

We begin by discussing our simulation setup and strategy for
exploring wing production in Section~2, then present our model runs and
describe the evolution of a generic winged source in Section~3.  In particular,
we find that favorable pressure gradients are necessary but not sufficient
to produce an X-shaped source; the character and time evolution of the jet is
equally important.  In Section~4, we briefly discuss the missing and 
artificial physics in our simulations, and in Section~5 we assess the
backflow model in view of our results as well as discuss the implications for
the broader picture of radio lobe morphology.  Finally, in Section~6 we
summarize our main results and conclusions.  

Throughout this paper
we use the terms ``secondary lobes'' and ``wings'' interchangeably.  
We also use the term ``winged'' source to refer to any radio source with 
substantial symmetric off-axis distortions whereas ``X-shaped'' sources are
a subset of winged sources with an axial wing-to-lobe length ratio of more
than $0.8$.  This nomenclature reflects the view of the hydrodynamic model in 
which XRGs are indeed a subset of a broader category of distortions.  Finally,
we use the term ``back-flows'' to refer specifically to actual fluid flows
heading back to the nucleus from the jet heads, whereas we use ``backflow'' to
refer to the material in these flows.  

\section{Hydrodynamic Simulations}

We use a parallelized version of the {\sc ZEUS}
code \citep{stone92a,stone92b} for our hydrodynamic simulations.
{\sc ZEUS} is a second-order (spatial accuracy) Eulerian finite-differencing
magnetohydrodynamic (MHD) code which solves the standard equations of hydrodynamics,
\begin{eqnarray}
\frac{\partial \rho}{\partial t} + \nabla \cdot (\rho {\bf v}) &=& 0\\
\frac{\partial {\bf v}}{\partial t} + {\bf v} \cdot \nabla {\bf v} &=& - \frac{\nabla P}{\rho} - \nabla \Phi\\
\frac{\partial}{\partial t} (\rho e) + \nabla \cdot (\rho {\bf v} e) &=& -P \nabla \cdot {\bf v},
\end{eqnarray}
for an ideal compressible fluid and introduces artificial viscosity for shocks.
The version we use, {\sc ZEUS-MPv2}, is based on
the National Center for Supercomputing Applications (NCSA) code
described in \citet{hayes06}.  
We use spherical polar coordinates ($r$,$\theta$,$\phi$) in the purely
hydrodynamical mode for all our runs; the jets
are therefore injected from an inner boundary sphere with some small, but
finite, radius $r_{\text{inner}}$.
We outline our basic simulation setup
below, followed by our strategy for exploring winged sources and a description
of the evolution of a standard double-lobed source for comparison.

\subsection{Simulation Setup}

Hydrodynamic and MHD simulations of radio galaxies 
interacting with their surroundings are common due to the possibility that 
energy deposited by the lobes is a mode of heating in AGN feedback scenarios.  
We do not incorporate relativistic jet physics since we are primarily 
concerned with lobe mixing and evolution, but the robustness of this assumption
is explored in Section~4.1.  We note here that non-relativistic, light,
hypersonic hydrodynamic jets reproduce several essential features of jets.
These include recollimation shocks,
terminal shocks at the jet head (associated with radio ``hot spots'') and 
back-flows of spent material from the terminal shocks which sheath the jet
and produce lobes.  Since we are interested in the evolution of back-flowing
fluid in anisotropic environments, we adopt a non-relativistic purely 
hydrodynamic scheme.

We inject the jets as bi-directional flows into
an initially hydrostatic, ellipsoidal atmosphere.  
The atmospheres we set up are ellipsoids with a polytropic equation 
of state ($\gamma = 5/3$) and a 3D $\beta$-model density profile 
\citep{cavaliere76} given by
\begin{equation}
\rho = \frac{\rho_0}{(1+(r/r_0)^2)^{3\beta/2}},
\end{equation}
where $\beta = 0.5$, $r_0$ is the core radius, and $\rho_0$ is the core
density.  To adapt this model for elliptical atmospheres, we make the radial
density profile dependent on angle by adding a major-axis position angle PA and
ellipticity $\epsilon$ to the atmosphere, which we denote as $r_e$ for an
elliptical atmosphere.  To easily allow for triaxial
atmospheres, we define the ellipsoid in Cartesian space ($x_e$, $y_e$, $z_e$)
and break up $\epsilon$ along each axis, then transform back to spherical 
coordinates to obtain the radius $r_e$ using the grid coordinates 
($r$, $\theta$, $\phi$):
\begin{eqnarray}
r_e & = & \frac{\sqrt{x_e^2 (1 - \epsilon_x)^2 + y_e^2 (1 - \epsilon_y)^2+
 z_e^2 (1-\epsilon_z)^2}}{(1-\epsilon_{\text{max}})} \\
x_e & = & r \sin(\theta)\biggl[\cos(\phi)cos(\text{PA})+\sin(\phi)\sin(\text{PA})\biggr]\\
y_e & = & r \sin(\theta)\biggl[\sin(\phi)cos(\text{PA})-\cos(\phi)\sin(\text{PA})\biggr]\\
z_e & = & r \cos(\theta)\biggl[\cos(\text{PA}) + \sin(\text{PA})\biggr],
\end{eqnarray}
where PA is nominally measured clockwise from the $z$-axis (in practice, it is
only the difference between the jet axis and the major axis of the ellipsoid
that matters).  To normalize the size of the atmosphere, $\epsilon_{\text{max}}$ in
Equation~5 is defined as the largest value of
$\epsilon$ along each axis.  For example, an atmosphere with large
$\epsilon_z$ is elongated along the $z$-axis.  
Equations~4 and 5 may be understood as 
saying that introducing $\epsilon$ changes the effective core radius along
a given axis: 
$r_{0,\text{eff}} = r_0 (1-\epsilon_{\text{max}})/(1-\epsilon)$.
For example, if $\epsilon_{\text{max}} = \epsilon_z = 0.75$, 
$\epsilon_x = 0.0$, and $r_0 = 1.0$, then along the $z$-axis $r_{0,\text{eff}} = 1.0$
and along the $x$-axis $r_{0,\text{eff}} = 0.25$.  This phrasing is convenient
when comparing the major and minor axes rather than the quantities along the
full range of angles $\theta$ and $\phi$ and we use it hereafter.

The atmospheres are initially set up in hydrostatic equilibrium, 
\begin{equation}
\nabla \Phi = \frac{1}{\rho} \nabla p,
\end{equation}
assuming that the background dark matter potential is static and dominant so that
the gas self-gravity is not important.  We define the adiabatic sound speed 
$c_s \equiv 1.0$ in code units and set up an isothermal atmosphere so the 
potential $\Phi$ is
\begin{equation}
\Phi = \frac{c_s^2}{\gamma} \ln(\rho)
\end{equation}
Since XRGs are powerful radio galaxies and exist at low redshift, we assume
a smooth intergalactic/intracluster medium (IGM/ICM) and gas-poor systems,
i.e. no disk of colder (i.e. atomic or molecular) material in the host 
\citep[c.f.][]{sutherland07}.  We note that in several exploratory runs, 
additional atmospheric complexity is overlaid onto a smooth atmosphere with
no change in the gravitational potential $\Phi$, i.e. not initially in 
hydrostatic equilibrium. 

The jets in our simulations are hypersonic ($\sim$100$c_s$) light
($\rho_{\text{jet}} \sim 0.01 \rho_0$) flows injected in pressure
equilibrium with the ambient material from back-to-back circular footpoints 
on the small inner boundary sphere at the origin.  
The kinetic luminosity of the jets is given by
\begin{equation}
L_{kin} \sim \tfrac{1}{2}\rho_{\text{jet}} v_{\text{jet}}^3 A_{\text{jet}}
\end{equation}
where $A_{\text{jet}}$ is the area of the footpoint at $r_{\text{inner}}$, 
and ignoring higher order contributions from the gravitational energy or 
thermal flux.  To tune $L_{\text{kin}}$ we primarily vary 
$v_{\text{jet}}$ because (i) $L_{\text{kin}}$ is most sensitive to changes in
$v_{\text{jet}}$, (ii) the jets must be ``light'' to ensure that the 
Kelvin-Helmholtz instability growth rate at the boundary of the cocoon is
approximately the same as in the relativistic case \citep{reynolds02}, and 
(iii) jets are highly
collimated.  The maximum jet width is constrained by high-resolution X-ray
observations of jet knots \citep[e.g.][]{perlman05} and very long baseline 
interferometry observations of transverse structures \citep[e.g.][]{gabuzda04}.
However, we note that in Section~3.2 we vary $\rho_{\text{jet}}$ and 
$A_{\text{jet}}$ in a limited range.  

We insist that the jet cover enough grid zones ($\sim 30$) to resolve 
transverse structures such as the oblique shocks which collimate the jet.  
The jet is rapidly precessed 
around a small angle ($\alpha < \theta_{\text{jet}}$) at $20\pi$ rad s$^{-1}$ 
(code time) to break the axisymmetry of the simulation setup and simulate 
helical instabilities, thus spreading the thrust out over a larger working 
surface \citep[c.f.][]{vernaleo06,heinz06,sutherland07,oneill10}.  The location 
of the jet on the inflow sphere is thus given by 
$(\theta,\phi) = (\alpha, \Omega_{\text{jet}} \cdot t)$.  Although the
angular quantities are free parameters, they are constrained by the observed
collimation of jets.  Typical values are $\theta_{\text{jet}} = \pi/7.5$ and
$\alpha = \pi/60$ for $r_{\text{inner}} = 0.05$.  In agreement with other work,
these jets develop a cylindrical core of fast-moving material sheathed in a
slower concentric shell continuous at the boundaries with the surrounding
material and core.  

We use a grid with zones spaced according to a geometric series in $r$ and $\theta$ (256 
bins each) and uniformly in $\phi$ (64 bins).  This grid 
resolves the internal jet structure and important processes near the
injection footpoints.  We choose $r \in [0.05,5.0]$ and 
$\theta \in [0.01, 3.13]$ to avoid the polar singularity and begin with
reasonably sized grid zones.  The $\theta$ grid is broken into two symmetric 
128-bin components with the smallest zones near the poles where the jets are 
injected.  The adequacy of our grid is demonstrated by a run with double the
resolution in each direction which produces similar internal jet structure.
Mixing is not substantially different at the lobe boundaries.  

We use periodic boundary conditions in the $\phi$ direction and reflecting 
boundary conditions in $\theta$.  The outer $r$ boundary at $r = 5.0$ has 
outflow conditions (material leaving the grid); likewise, outflow conditions
exist for all $r_{\text{inner}}$ zones except where the jets are injected.
The outer boundary is far from the jet activity so negligible material leaves the 
grid there.  We discuss the importance of the inner boundary sphere in Section~4. 

Code units are transformed to physical units by choosing appropriate
values for $r_0$, $c_s$, and $\rho_0$.  For instance, \citet{reynolds02}
defined $r_0 = 100$~kpc, $c_s = 1000$~km~s$^{-1}$ and 
$\rho_0 = 0.01 m_{\text{H}}$~g~cm$^{-3}$ for a rich cluster and
$r_0 = 10$~kpc, $c_s = 500$~km~s$^{-1}$ and 
$\rho_0 = 0.1m_{\text{H}}$~g~cm$^{-3}$ for a group or poor cluster.  In
the former scheme a code unit of time (derived from the crossing time) 
corresponds to 50~Myr and in the latter 10~Myr.  In our runs, we vary $r_0$
and $\rho_0$ but fix $c_s$; a value of $c_s \sim 500$~km~s$^{-1}$ appears 
appropriate for XRGs based on temperatures derived from spectral fitting
\citep{hodges10a,landt10}.  A jet injected at $100c_s$ would then have a
physical speed between $0.17c - 0.35c$ (a Lorentz factor of $\gamma \sim 1.01 - 1.05$).  
However, as noted by \citet{kom96},
comparing classical and relativistic jet simulations requires careful matching
of parameters, in particular the mass-energy density content of the jet.  
Hence, the jet velocities chosen should not be taken directly as assumptions of
true jet velocity.

Our runs were parallelized and used variety of processors.  Many of the
runs were conducted on quad-core Intel$^{\text{\textregistered}}$ 
Core$^{\text{\texttrademark}}$ $2.4$ and $2.83$~GHz workstations.

\subsection{Strategy}

We now outline our guiding strategy to determine whether X-shaped sources can 
result from the interaction of radio galaxy lobes with their environment.

First, we only model powerful FR~II 
sources which produce strong back-flows.  Neither the physical origin of the
FR~I/II dichotomy nor the differences between jets in radio loud and radio
quiet sources are understood, but the higher luminosity FR~IIs exhibit the
hot spots, well defined lobes, and bridges exploited by the backflow model.

In accordance with the observations \citep{capetti02,saripalli09,hodges10a}, 
we inject jets along the major axis of the surrounding atmosphere. 
We first use unrealistic
atmospheres with very favorable pressure gradients for wing production
\citep[as in][]{capetti02}, then explore jet and atmosphere parameters
to study the production and characteristics of XRGs.  In particular, we 
explore the dependence of wing formation on jet width, density, and
kinetic luminosity as a function of time, and initial atmosphere parameters
(core radius, density, ellipticity, and position angle).  Although some 
parameter combinations are degenerate, this is a very large parameter space
because the $L_{\text{kin}}(t)$ may include dead time and intermittency.
Compared to 2D modeling, our 3D simulations eliminate the requirement of
axisymmetry which enhances the jet head advance speed \citep{bodo98} and 
the coherence of the back-flows.  The effect of turbulence is also more
realistically explored in three dimensions.

Motivated by the expectation that pressure gradients affect the wings
fundamentally the same way in different systems, we start from the ansatz that
jet and atmospheric parameters are orthogonal.  In other words, the jet
parameters may be tuned in some fiducial atmosphere and the atmosphere 
parameters may be tuned with some fiducial jet such that the behavior of
an arbitrary jet in an arbitrary atmosphere can be inferred. 
On the basis of this method we will evaluate
the factors key to wing prominence and the viability of the backflow models.

\begin{center}
\begin{figure*}[t] 
\centerline{
\includegraphics[scale=0.75]{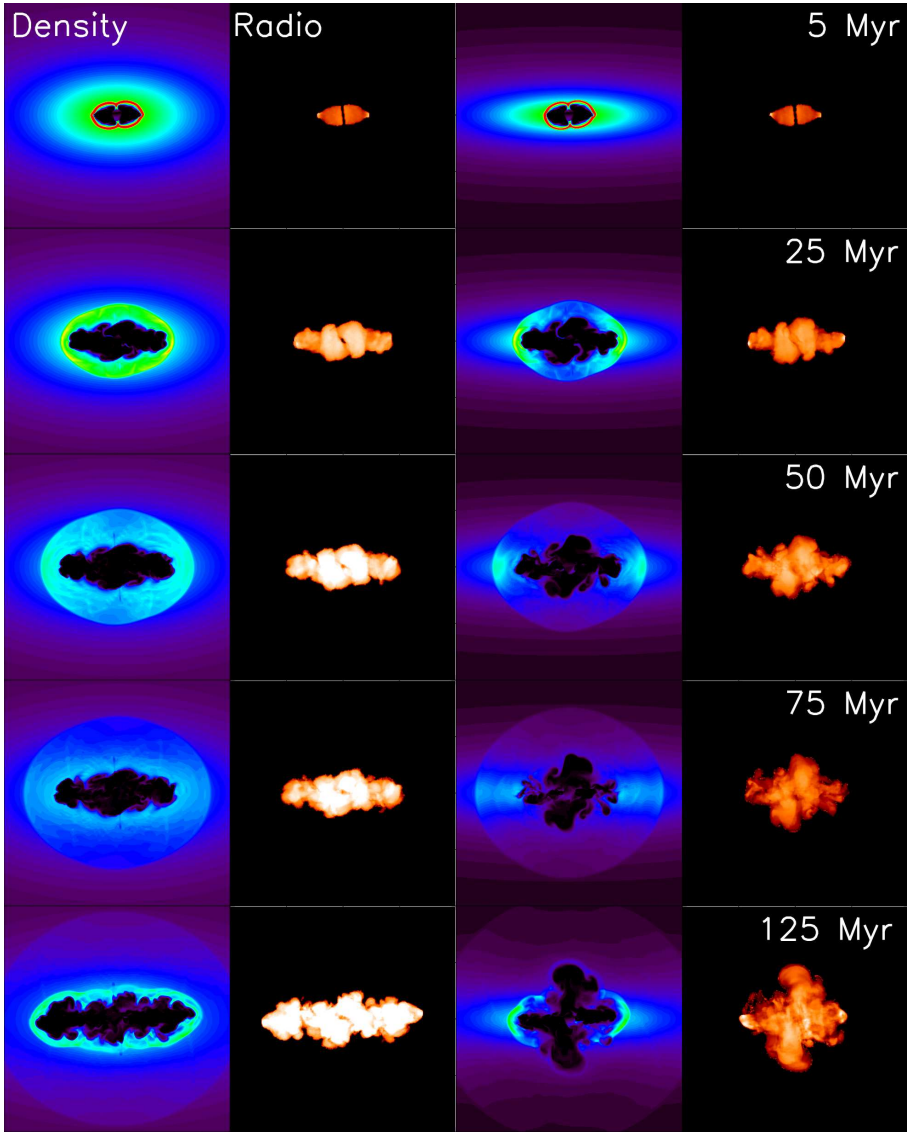}
}
\figcaption{{\small Comparison of a normal (left panels) and winged (right panels) radio
source using the same jet ($v_{\text{jet}} = 100\exp[-3t]$) 
in atmospheres differentiated only by $\epsilon$.
Note the size of the overpressured cocoon relative to the size of the 
atmosphere in the top panels.  At a code time of 1.0 (100~Myr) the jet
was restarted ($v_{\text{jet}} = 100\exp[-3(t-1.0)]$).  The colorbars are the
same in each image; the integrated radio is brighter for the normal source
because the winged source has the same amount of material spread out over
more volume.}}
\label{coevol}
\end{figure*}
\end{center}

\subsection{Hydrodynamic Models of Powerful Double-Lobed Radio Galaxies}

If wings are distortions to a generic double-lobed source, we expect our 
simulated radio galaxies to strongly resemble the double-lobed sources produced 
by earlier models.  Hydrodynamic and MHD models of jets are commonly employed 
to study either the phenomenology of the jets and lobes (with relativistic and
nonrelativistic fluids) or the energy deposited by the lobes into their 
environment.  Hence, a large body of work concerned
with phenomenological models of these sources exists and has produced key
insights into the life of a powerful radio source.

These studies include work on early jet/lobe evolution in a relaxed atmosphere
\citep[e.g.][]{krause05,antonuccio10}, the influence of the 
$\rho_{\text{jet}}/\rho_0$ density contrast and 
internal Mach number \citep[e.g.][]{carvalho02a,krause03,vernaleo07}, jet stability 
in nonrelativistic and relativistic conditions 
\citep[e.g.][]{rosen99,keppens08,oneill10,mignone10}, 
the importance of intermittency \citep{oneill10} or intrinsically spreading out
the jet thrust to slow the advance of simulated jets \citep[e.g.][]{heinz06},
the importance of the jet environment to morphology or energy deposition
\citep[e.g.][]{capetti02,carvalho02b,krause05,zier05,vernaleo07,simionescu09,
kawakatu09,falceta10,gaibler10}, and physics beyond MHD \citep[e.g.][]{saxton10}.  
Naturally, there is
overlap between the phenomenological studies and those motivated by
the challenge to produce jets which prevent a cooling catastrophe in the ICM
of a host galaxy cluster by 
isotropizing energy distribution.  
The viability of radio galaxies as AGN feedback
mechanisms is beyond the scope of this work, but jet lifetime and the passive
evolution of ``dead'' radio galaxies \citep{reynolds02} place important 
constraints on wing prominence.  

Based on this work, we understand a powerful double-lobed source to have
three distinct phases of evolution: (i) \textit{ignition}, in which a spheroidal cocoon of 
relativistic plasma is formed around the nascent jets, (ii) the \textit{active phase},
wherein the jet produces the cigar-shaped lobes associated with FR~II radio
galaxies, and (iii) the \textit{passive phase} where lobes evolve after the jet is
turned off.  Below we outline these phases for a preliminary hydrodynamical 
simulation of a fast ($v_{\text{jet}} = 100c_s \exp(-3t)$), light jet with large $L_{\text{kin}}$ (a ``FR~II'' 
source) to provide a framework for discussing the evolution of winged 
sources (left-hand panels of Figure~\ref{coevol}).  We note that 
\citet{belan11} have recently found good agreement between the structures 
observed in hydrodynamic simulations of hypersonic jets and laboratory 
experiment.

\paragraph{{\it Ignition}}
The jet is injected in pressure equilibrium with the initially relaxed 
surrounding medium.  The flow quickly forms and drives a spheroidal bow shock
into the surrounding medium (Figure~\ref{coevol}, top left panel).  At the
location where the jet impacts the shock, back-flows develop which fill the
space evacuated by the bow shock with light, hot spent jet plasma.  This
plasma forms a cocoon which sheaths the jet.  At very early times the
jet expands laterally, since it is unconfined by the cocoon plasma (in part due
to the initially conical shape of the jet) and the expansion of the nascent
radio galaxy is nearly self-similar \citep{carvalho02a,krause05}.  However,
as a result, the early back-flows acquire a circulatory motion and flow along
the inner edge of the bow shock (Figure~\ref{backflow}, top panel).  By the
time these flows reach the midplane between the two jets, they have velocity
vectors pointing radially inwards and do not collide with back-flows from
the counter-jet \citep{antonuccio10}.  In our models, the back-flows are
also prevented from interacting with their counterparts from the counter-jet
because our jets are injected into pristine atmospheres such that at very
early times, each jet inflates its own cocoon.  The two cocoons are both
bounded by a bow shock, and the two shocks meet at the midplane between the
jets and form an interstice of dense material that prevents early mixing
(Figure~\ref{backflow}, top panel).  Whether such an interstice is present
in real sources is not clear \citep[we do not resolve the very earliest jet
stages such as the ``flood-and-channel'' phase seen in a clumpy, warm
disk in][]{sutherland07}, but the interstice is ablated and the cocoon is
unified by the active phase. 

As the cocoon pressure builds due to backflow confined by the bow shocks,
the jet becomes azimuthally confined and a recollimation shock appears near
the injection point; by the end of the ignition phase, the cocoon is 
strongly overpressured relative to its environment.  At the same time, 
the terminal shock becomes increasingly distinct from the bow shock, and the 
jet head takes on the characteristic double-pronged appearance of a 3D 
hydrodynamical jet.  While the back-flows still follow the bow shocks during 
the late ignition phase, they are increasingly straight.  We note that this
early evolution is seen in all our runs and does not depend much on atmosphere
or jet parameters, nor on the boundary conditions (as long as the core radius
of the atmosphere $r_0$ is much larger than $r_{\text{inner}}$). 

\paragraph{{\it Active Phase}}
When the jet head overtakes and pierces the initial bow shocks
\citep[driving a bow shock contiguous with the initial burst;][]{krause05},
the active phase begins and a classical radio galaxy develops, with a 
cigar-shaped cocoon, hot spots, bow shocks (middle-left panels of Figure~\ref{coevol}), 
and straight back-flows (Figure~\ref{backflow}, bottom panel). 
At this point, the cocoon ceases to be substantially overpressured and the
cocoon's lateral expansion falls close to the sound speed of the ambient
medium, lagging behind the bow shock (the cocoon expands due to buoyancy, 
Kelvin-Helmholtz, and Rayleigh-Taylor instabilities).  
This weak shock ($v_{\text{shock}}
\sim 1.5c_s$) sweeps up a large amount of material as it expands and becomes
better described as a strong sound wave (Figure~\ref{coevol}).  Meanwhile,
the jet develops internal oblique collimating shocks along its length,
retaining a high velocity out to the terminal shock at the jet head 
(i.e. the angular size of the jet decreases with increasing radius) where the
cocoon becomes momentum-driven.  Less powerful jets
do not produce cocoons \citep{vernaleo07}, but we focus on the cocoon-bounded
case \citep[the cocoon pressure also depends on the internal Mach number;][]{
carvalho02a}.  

\paragraph{{\it Passive Phase}}
Once the jets are turned off, the back-flows cease and the ``dead''
radio galaxy rises buoyantly in the atmosphere, lifting large amounts of
material to great heights and mixing with the surrounding medium
\citep{reynolds02}.  Although the
radio galaxy is relatively efficient at depositing its energy irreversibly in
the surrounding medium, it is difficult to isotropize this energy deposition
on short enough timescales to avoid a cooling catastrophes 
\citep[e.g.][]{reynolds02,omma04a,omma04b,vernaleo06,vernaleo07,oneill05,deyoung10,
oneill10,ostriker10}.  The cocoon separates into bubbles which pinch
off along the direction of the jets and rise in opposite directions; the
evolution of these bubbles may be quite complex \citep[e.g.][]{begelman01,churazov01,
ruszkowski07,dong09,oneill09,braithwaite10,pope10}.  This very late stage
evolution is unlikely to be important for winged galaxy evolution, as winged
galaxies appear to be related to sources with strong bridges \citep{leahy84},
and is not shown in Figure~\ref{coevol}.

\section{Results}

With the scheme described in Section~2, we produce winged radio galaxies
by the deflection of backflow into channels perpendicular to the jets. 
In our models, wings are produced in strongly asymmetric atmospheres when
the jet is driven near the major axis; the wings are produced along the
minor axis as cocoon material escapes the central regions of the atmosphere
and evolve buoyantly, collimated by the surrounding stratified atmosphere
which promotes expansion along the steepest pressure gradient.  Hence, these
wings are similar to those produced by \citet{capetti02} and \citet{zanni03}.
Without introducing additional complexity to the atmosphere, hydrodynamic wings
are formed in two stages.  First, during the ignition phase (Section~2.3), 
the overpressured cocoon expands faster along the minor axis due to the 
steeper pressure gradient and forms small channels (proto-wings) 
perpendicular to the jet into which back-flowing plasma flows.  The proto-wings
produced in this way can account for 20--40\% of wing length at the end of the
active phase depending on the gradient.  Second, during the active phase, the
wings rise buoyantly, fed by turbulent flows near the midplane.  Although the
back-flows near the jet heads are initially laminar and supersonic (relative
to the lobe material), they quickly decelerate and do not enter the wings as
coherent flows; the wings expand subsonically.  

During the active phase the wings evolve almost independently of
the cocoon.  Hence, once wings have developed, their length depends only on
the properties of the native atmosphere whereas the length of the jet-fed
primary lobes is dominated by the properties of the jet (in particular the
kinetic luminosity as a function of time) for powerful jets.  Therefore, 
\textit{decaying} jets produce prominent wings. 

In this section, we explore these ideas in detail, first comparing the life of
a winged source to a canonical double-lobed one (Figure~\ref{coevol}) and 
then looking at the dependence of wing prominence and longevity on various 
tunable parameters in our models.

\subsection{Evolution of a Winged Source}

Winged galaxies experience the same life stages outlined in 
Section~2.3.  We describe how wings fit into this process in detail below,
referencing our standard (extremal) atmosphere (Run~{\tt STANDARD}, 
Tables~\ref{atmostable} and \ref{jettable}).  While unrealistic, this atmosphere provides
important insights into the backflow model and can be directly compared to
prior work \citep{capetti02,zanni03}.

\paragraph{{\it Ignition}}
During the ignition phase of a winged source, the anisotropic expansion of the 
overpressured cocoon in an aspherical atmosphere produces channels which
will later become the wings (the proto-wings).  This pressure-driven expansion
is supersonic but brief, since the cocoon quickly reaches pressure 
equilibrium.  Although these channels need not be produced by an overpressured
cocoon, some channels must exist for wings to form.  

The degree of cocoon expansion in a given direction depends on the pressure
gradient experienced in that direction, and hence on the atmospheric
parameters.  In particular, the eccentricity of the atmosphere effectively
changes the core radius $r_0$ seen by the cocoon in different directions.
For example, along the $x$-axis ($r\hat{x}$; the jet is driven along 
$\hat{z}$) Equation~4 gives:
\begin{equation}
\frac{\partial p}{\partial x} = -\frac{3}{2} \frac{c_s^2}{\gamma} \rho_0 r_{0,\text{eff}}^{3/2} \frac{x}{(x^2 + r_{0,\text{eff}}^2)^{7/4}}.
\end{equation}
where $r_{0,\text{eff}} = r_0 (1-\epsilon_{\text{max}})/(1-\epsilon_x)$ as 
above.  Since the pressure gradient is steeper for smaller $r_{0,\text{eff}}$, 
higher ellipticity along other axes promotes wing expansion:
for $\epsilon_x = 0.0$ and $\epsilon_z = \epsilon_{\text{max}} = 0.75$,
$r_{0,\text{eff}} = 0.25 r_0$ along the $x$-axis.  Conversely, if 
$\epsilon_x = \epsilon_{\text{max}} = 0.75$ and $\epsilon_z = 0.0$ (the jet
is pointed along the minor axis), $r_{0,\text{eff}} = r_0$ and wings are 
suppressed.  Of course, wing expansion depends on the actual pressure
gradients rather than just the ratio along different axes; the base values
$r_0$ and $\rho_0$ determine whether wings can form (i.e. a highly elliptical
atmosphere can have shallow pressure gradients along the minor axis if it is
very large).  Note that at very small radii $r \ll r_{0,\text{eff}}$ 
(i.e. during the ignition phase), the pressure gradient steepens linearly
with increasing $r$ and at large radii (during the active phase, see below) 
the pressure gradient becomes shallower as $r^{-1/2}$.  Hence, the ignition
phase is the time at which atmospheric asymmetry has the strongest effect on
the ultimate morphology of the source.  

\begin{figure}[t]
\begin{center}
\includegraphics[scale=0.40]{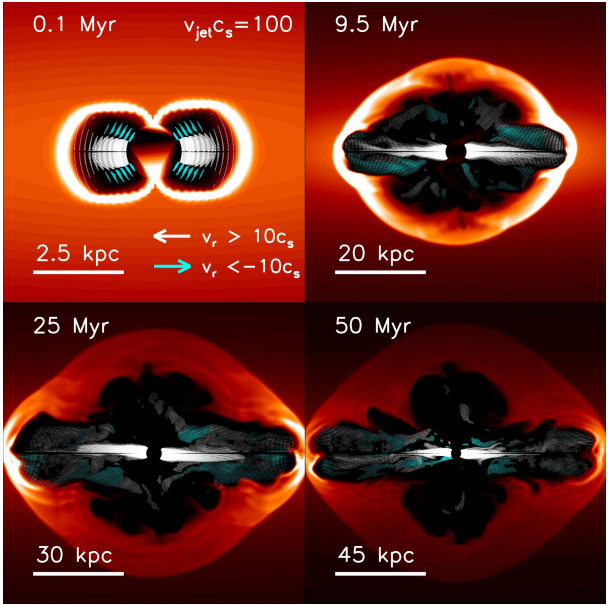}
\caption{{\small We show here $v_x$ and $v_y$ velocity vectors overlaid on density
slices (taken at $\phi = 0$) at different times for the simulation TX\_E75\_VE3\_B5.0 (Table~3).  
We have chosen (arbitrarily) $c_s = 500$~km~s$^{-1}$ and $r_0 = 50$~kpc 
(corresponding to a code $r_0 = 2.0$) to represent a large elliptical galaxy.
White
vectors point radially outwards and cyan vectors point radially inwards (i.e.
the back-flows).  Vectors are only shown when the magnitude of $v$ exceeds
the sound speed of the lobe material ($c_{s,\text{lobe}} = 10c_s$).  Note
that early on, the back-flows follow the contact discontinuity and are
directed towards the inner boundary, but later become straight as in
\citet{antonuccio10}. 
}
}
\label{backflow}
\end{center}
\end{figure}

\begin{figure}
\begin{center}
\includegraphics[scale=0.35,angle=90]{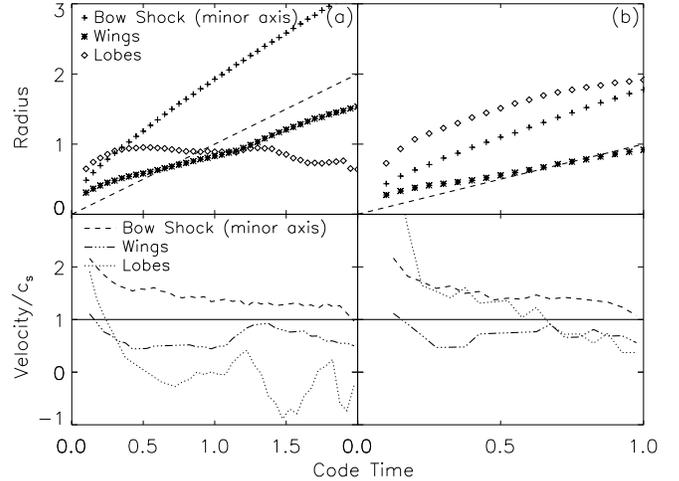}
\caption{{\small Positions and velocities of the leading edges of the bow shock,
wings, and lobes shown for (a) the standard simulation and (b) a triaxial
simulation.  The dashed line in the top panel shows the position of a 
point moving radially outwards at the sound speed.  Note the different
$x$-axes; at $t = 1.0$ in the left panels, a new jet is reinjected.  
\textit{Negative} growth for the lobes indicates collapse during the passive
phase.}}
\label{wingvel}
\end{center}
\end{figure}

\paragraph{{\it Active Phase}}
Once the cocoon has reached pressure equilibrium, the proto-wings are no
longer pressure-driven.  As the back-flows become straight, they merge and
fill the channels (Figure~\ref{backflow}, bottom panel) to form structures 
more closely resembling observed wings.  However, the flows into the wings are
turbulent and transonic or subsonic (relative to the internal lobe sound speed 
$c_{s,\text{lobe}} = 10c_s$).  In other words, wings are not \textit{driven}
during the active phase.  Rather, because they are structures filled with
light fluid, they rise buoyantly in the atmosphere, collimated by the 
stratified elliptical atmosphere.

Since the wings rise buoyantly, their growth rate is subsonic.  To see this,
it is instructive to look at the simplified case
of a spherical bubble (of fixed radius) rising buoyantly in a dense fluid.
The (terminal) buoyant velocity at a given height for such a bubble is
\begin{equation}
v_{\text{buoy}} = \sqrt{\frac{2g V_{\text{b}}}{C_D A_{\text{b}}}} =
c_s \sqrt{\frac{16}{3\gamma} \frac{r_{\text{b}}}{r_{0,\text{eff}}^2} \frac{r}{1+(r/r_{0,\text{eff}})^2}} 
\end{equation}
where $V_{\text{b}}$, $A_{\text{b}}$, and $r_{\text{b}}$ are the
volume, cross-sectional area, and radius of the bubble respectively, $C_D = 0.75$
is the ``drag'' coefficient, and we have used the background dark matter
potential $\Phi$ to compute $g$ in terms of $r$ and effective core radius
$r_{0,\text{eff}} = r_0 (1-\epsilon_{\text{max}})/(1-\epsilon)$.  Naturally, 
this follows the pressure gradient (Equation~12), so when the bubble is at
very small radii its velocity \textit{increases} as $r^{1/2}$, whereas outside
$r_{0,\text{eff}}$ it \textit{decreases} as $r^{-1/2}$.  For typical values in
our simulations, the peak value of $v_{\text{buoy}}$ is approximately $c_s$.
Hence, in a highly elliptical atmosphere, a bubble rising along the minor axis
will rise subsonically with decreasing velocity for most of its lifetime.

Our wings do not fit this simplified case because they are large relative to
$r_{0,\text{eff}}$, variable in size, aspherical, and connected to the
cocoon.  Nonetheless, their expansion is subsonic for the same reason: the
wings exceed $r_{0,\text{eff}}$ early on when their growth is pressure-driven
rather than buoyant, and $v_{\text{buoy}}$ monotonically declines thereafter.
This transition from supersonic to subsonic wing expansion is plainly seen in
Figure~\ref{wingvel}.

\begin{figure}[t]
\begin{center}
\centerline{\includegraphics[scale=0.35,angle=90]{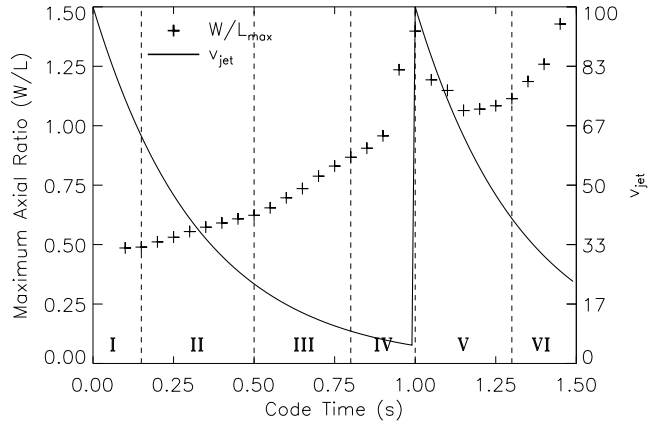}}
\caption{{\small Plot of $W/L_{\text{max}}$ (crosses) in timestep intervals of $\delta t = 0.05$
for the standard simulation with a reinjection at $t = 1.0$.  We have
overplotted the velocity of the jet as the black line (right $y$-axis).
The Roman numerals indicate (I) the overpressured cocoon phase, (II) the
active phase (powerful jet), (III) the active phase (weak jet), (IV) the
passive phase (with cocoon collapse), (V) a second active phase due to the
reinjected jet, and (VI) the second passive phase with further cocoon
collapse.  During cocoon collapse, the source does not resemble a winged
source (instead it is a ``dead'' radio galaxy).  A source appears winged
during the active phase when the jet is most like a weak FR~II or strong
FR~I.}}
\label{standardgrowth}
\end{center}
\end{figure}

\begin{figure}[b]
\begin{center}
\includegraphics[scale=0.54]{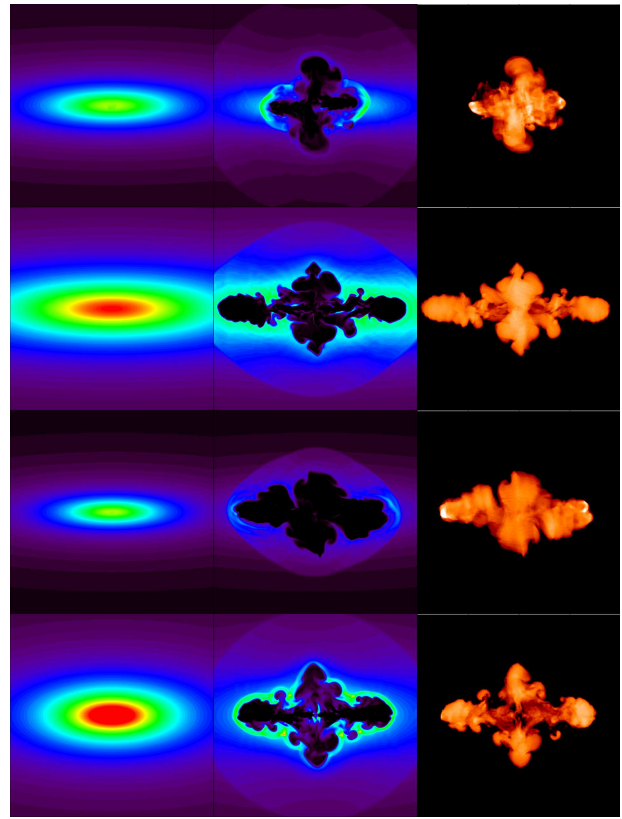}
\caption{{\small Examples of different intrinsic wing morphology.  Left: Initial
relaxed atmosphere density.  Center: Winged sources during the late active
phase.  Right: False synchrotron maps ($j_{\nu} \propto p/\rho^{7/4}$
assuming equipartition) of these sources.}}
\label{diffwings}
\end{center}
\end{figure}

Therefore, wings in our models may not intrinsically exceed the length of
the primary lobes unless the primary lobes advance at some average speed 
$\bar{v}_h < c_s$.  Since powerful jets are required to produce the proto-wings
in an overpressured cocoon phase, prominent wings require decaying jets,
intermittent jets, or jets which deposit their thrust over an increasingly
large area with time.  The profile of the fiducial run
is shown along with its wing prominence in Figure~\ref{standardgrowth} and
compared to a triaxial atmosphere in Figure~\ref{wingvel}.  Note that in 
the fiducial run (left panel of Figure~\ref{wingvel}) the wings eventually
overtake the lobes (in part due to the collapse of the lobes when the
jet is very weak), but clearly move into a subsonic regime early and remain
there.  The bow shock remains mildly supersonic through most of both 
simulations.  In Figure~\ref{diffwings} we show wings produced by decaying
jets during the active phase in several environments (the fiducial run is
shown in the top row).

\paragraph{{\it Passive Phase}}
Once the jets are turned off, the cocoon disintegrates as the lobes either
rise buoyantly as bubbles or collapse under
the relaxing atmosphere.  If the radio galaxy is within a few $r_0$ in a dense
atmosphere, the fall-back of displaced material shreds the lobes into small
bubbles.  In either case, the wings pinch off and rise.  These bubbles do not
survive long, but their behavior might be substantially altered in the 
presence of magnetic fields \citep[e.g.][]{oneill05,braithwaite10,pope10}.  
Because observed winged sources have strong bridges, we expect them to be
in the active phase.

\paragraph{{\it Reinjection}}
A powerful jet reignited during the passive phase before the cocoon has disintegrated
may significantly enhance the wings.  If the old jet channel is somewhat
broken up, the reinjected jet forms a new terminal shock and bow shocks \textit{inside} the
old cocoon.  These new shocks do not form spheroidal structures (as in a 
relaxed atmosphere) but instead produce strong, straight back-flows near 
the midplane.  Hence, the wings receive a large influx of fresh supersonic
plasma directly after reinjection (Figure~\ref{coevol}, bottom panels).  This
brightens the wings substantially and reinforces their structures.

There is a relatively narrow window of time where this process is effective.
If the reinjection occurs while the jet channels are largely intact (i.e. 
during the active phase), the jet simply follows these channels.  
On the other hand, if the reinjection occurs when the
wings have already separated from the lobes as individual bubbles, the new
jet cannot feed them.  Even if the reinjection occurs at the ``right'' time,
the efficacy of the restarted jet at promoting wings is short-lived.
However, this mechanism can produce wings that are intrinsically
\textit{longer} than the jet-driven lobes (Figure~3, bottom panels).  
If the reinjected jet decays,
this extreme axial ratio can be maintained for most of the lifetime of the
restarted radio galaxy and a bona-fide X-shaped source results, although this
source would only have the ``FR~II'' primary lobes for a short period of
time ($\sim 5$~Myr for a source 40~kpc across). 

\subsection{Ecology of Winged Sources}

\begin{figure}[t]
\begin{center}
\includegraphics[scale=0.55]{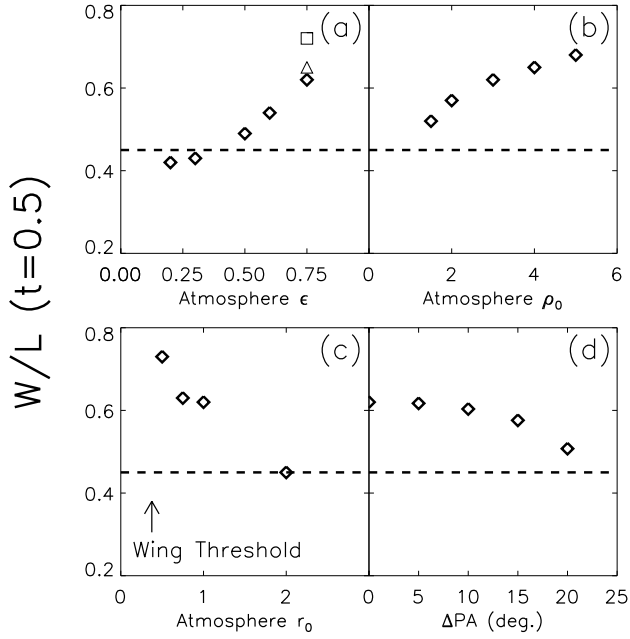}
\caption{{\small These plots show the dependence of axial ratio $W/L$ on (a) the
ellipticity of the atmosphere $\epsilon$, (b) the core density $\rho_0$, (c) the
core radius $r_0$, and (d) the angle between the jet and the major axis
$\Delta \text{PA}$.  Each plot represents varying one parameter while
holding the other standard atmosphere parameters ($\epsilon = 0.75$, 
$\rho_0 = 3.0$, $r_0 = 1.0$, and $\Delta \text{PA} = 0$) steady and using
the standard jet.  In panel (a) the square and triangle represent runs with 
different $\epsilon_x$ for $\epsilon_z = 0.75$, demonstrating the effect of
triaxiality on $W/L$ (Table~\ref{atmostable}).  $W/L$ is measured at
$t = 0.5$ for all cases; for the standard jet, this represents the transition
during the active phase from a powerful to a weak jet (this time and the
``wing threshold'' vary with choice of jet).  Values taken from Table~\ref{atmostable}.}}
\label{atmosplots}
\end{center}
\end{figure}

\begin{figure}[t]
\begin{center}
\includegraphics[scale=0.55]{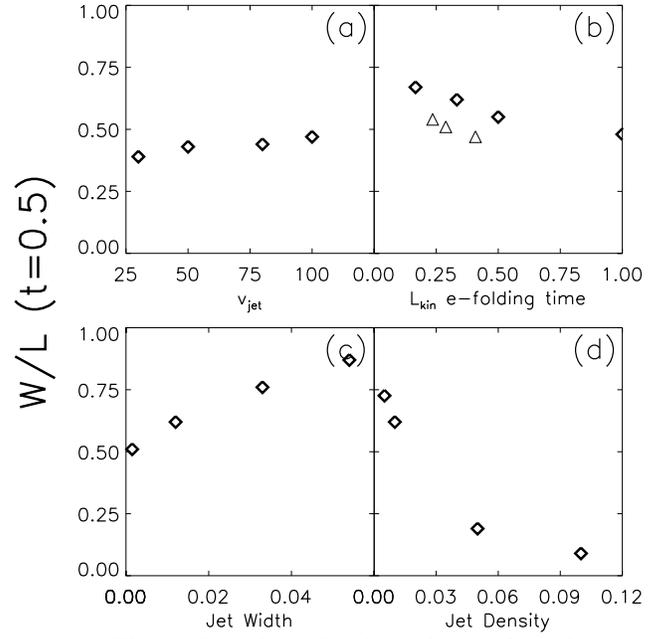}
\caption{{\small These plots show the dependence of axial ratio $W/L$ on (a) jet
velocity for a constant-velocity jet, (b) e-folding time for a decaying
jet (diamonds represent jets of the form $v_{\text{jet}} = 100\exp[-at]$ and
triangles $v_{\text{jet}} = 100\exp[-at^2]$), (c) area of the jet nozzle, and
(d) density of the jet material.  All runs are conducted in the standard
atmosphere (Table~\ref{atmostable}).  In (a) and (b), the jet kinetic luminosity
is \textit{not} conserved between runs, whereas in (c) and (d) we adjust 
the initial velocity $v_0$ to conserve $L_{\text{kin}}$.  As in
Figure~\ref{atmosplots}, we show $W/L$ as measured at a code time 
$t = 0.5$; all sources are in the active phase but not at the same place.
}}
\label{jetplots}
\end{center}
\end{figure}

In Section~3.1, we describe the life stages of a winged radio source.  We
now describe the dependence of wing prominence on our tunable parameters.
Following the strategy laid
out in Section~2.2, we begin with the fiducial simulation {\tt STANDARD} which 
is characterized by an atmosphere with a large ellipticity $\epsilon$ and
a jet with a velocity profile $v_{\text{jet}} = 100c_s\exp[-3t]$ (Figures~\ref{coevol}
and \ref{wingvel}).  
Taking the jet from the fiducial run, we individually vary parameters in the 
atmosphere in order to see their influence on wings (Figure~\ref{atmosplots},
Table~\ref{atmostable}), and then vary jet parameters in the fiducial atmosphere 
(Figure~\ref{jetplots}, Table~\ref{jettable}).  We then synthesize the 
information gleaned from these single-parameter curves to attempt to find 
atmospheres and jets which can explain real winged sources, including more 
complex behavior as well (Table~\ref{synthesistable}).  All of the runs
proceed to a code time of 1.0 or more, where most have entered the passive phase. 
Reinjections, where used, occur near the beginning of the passive phase.  

To quantify wing prominence we use the wing-to-lobe axial ratio ($W/L$).
This quantity is an imperfect measure because $W/L$ intrinsically varies with
azimuthal angle $\phi$ and time, and observable sources are seen in projection
so the observed $W/L$ will differ from the intrinsic value.  $W/L$ is also 
essentially meaningless during ignition or the passive phase, but a source
in one of these stages would not be classified as ``winged.''  For the
parameter exploration in Figures~\ref{atmosplots} and \ref{jetplots} (values in
Tables~\ref{atmostable} and \ref{jettable}), we adopt as a fiducial value
$W/L_{0.5}$: the maximum intrinsic $W/L$ for any pair of azimuthal angles
[$\phi$,$\phi+\pi$] at a code time $t = 0.5$.  For the standard jet 
($v_{\text{jet}} = 100c_s\exp[-3t]$), $t = 0.5$ represents the transition from a 
powerful to weak jet in the active phase, and by this time the wings have 
begun to grow buoyantly.  
Hence, a code time of $t = 0.5$ is a reasonable place to measure the influence
of the atmosphere on wings.  We also use $W/L_{0.5}$ in varying the jet 
parameters, noting that while $t = 0.5$ is no longer special, all of the runs 
in Table~\ref{jettable} are in the active phase at this time.  
$W/L_{0.5}$ is not predictive of wing length later in the same
simulation, but is a good measure of relative wing prominence between 
simulations due to the subsonic growth of wings.

From Figure~\ref{atmosplots} one can get a broad sense of the dependence of 
wing prominence on the size and shape of the atmosphere.  It is immediately
obvious that there is a strong dependence on the ellipticity $\epsilon$ of the
atmosphere, and that smaller, denser atmospheres are the most conducive to
wing formation (although ``small'' could be physically quite large depending on the
jet).  It is also notable that relatively high values of $W/L$
can be achieved by $t = 0.5$; because we use a decaying jet, the wings will
only become more prominent during the later active phase.  While these results
are not surprising (Equation~12), the particular form of the curves depends on
both the jet and the atmosphere.  As expected from \citet{capetti02} and
\citet{saripalli09}, long wings require jets co-aligned close to the major
axis.  

The trends are less clear when varying jet parameters (Figure~\ref{jetplots}).
In the top panels, the jet velocity is allowed to vary freely without conserving
integrated kinetic luminosity between runs.
If the jet is injected at constant velocity throughout the simulation
(Figure~\ref{jetplots}a), a faster jet is slightly better than a slower one at 
making wings.  However, a decaying jet (Figure~\ref{jetplots}b) is better
still, with pure exponential decay more effective at increasing $W/L_{0.5}$
than ``Gaussian'' jets of the form $v_{\text{jet}} = v_0 \exp[-at^2]$.  In
the bottom panels of Figure~\ref{jetplots}, we hold the kinetic luminosity
constant between runs by modifying $v_0$.  From Figure~\ref{jetplots}c, it is
evident that increasing the width of the jet (at the cost of a slower jet) is
important.  Changing the density of the jet material
has an extreme effect on $W/L_{0.5}$ (Figure~\ref{jetplots}d); above 
$\rho_{\text{jet}} \sim 0.01$, the jet ceases to be ``light'' relative
to the background and no longer forms lobes resembling radio galaxies
\citep{reynolds02}.  At very
low densities, mixing becomes very efficient at disrupting the radio galaxy.

We now discuss these results in detail, beginning with the properties of the
fiducial run.

\begin{deluxetable}{lccccccc}
\tablenum{1} 
\tabletypesize{\small}
\tablecaption{Varying the Atmosphere using the Standard Jet} 
\tablewidth{0pt} 
\tablehead{ 
\colhead{Name} & \colhead{Wings?} & \colhead{$r_0$} & \colhead{$\rho_0$} &
\colhead{$\epsilon_x$} & \colhead{$\epsilon_z$} & \colhead{$\Delta$PA} & 
\colhead{$W/L$} \\  &  &  &  &  &  & (deg.) & $(t = 0.5)$
}
\startdata
\multicolumn{8}{c}{Standard Atmosphere}\\
\cline{1-8}\\
STANDARD      & Y & 1.0 & 3.0 & 0.0  & 0.75 & 0.0  & 0.62 \\ 
\cline{1-8}\\
\multicolumn{8}{c}{Ellipticity}\\
\cline{1-8}\\
SJ\_E20       & N & 1.0 & 3.0 & 0.0  & 0.20 & 0.0  & 0.42 \\ 
SJ\_E30       & N & 1.0 & 3.0 & 0.0  & 0.30 & 0.0  & 0.43 \\ 
SJ\_E50       & N & 1.0 & 3.0 & 0.0  & 0.50 & 0.0  & 0.49 \\ 
SJ\_E60       & Y & 1.0 & 3.0 & 0.0  & 0.60 & 0.0  & 0.54 \\ 
\cline{1-8}\\
\multicolumn{8}{c}{Core Radius}\\
\cline{1-8}\\
SJ\_R0.5      & Y & 0.5 & 3.0 & 0.0  & 0.75 & 0.0  & 0.73 \\ 
SJ\_R0.75     & Y & 0.75& 3.0 & 0.0  & 0.75 & 0.0  & 0.61 \\ 
SJ\_R2.0      & N & 2.0 & 3.0 & 0.0  & 0.75 & 0.0  & 0.45 \\ 
\cline{1-8}\\
\multicolumn{8}{c}{Core Density}\\
\cline{1-8}\\
SJ\_D1.5      & Y & 1.0 & 1.5 & 0.0  & 0.75 & 0.0  & 0.52 \\ 
SJ\_D2.0      & Y & 1.0 & 2.0 & 0.0  & 0.75 & 0.0  & 0.57 \\ 
SJ\_D4.0      & Y & 1.0 & 4.0 & 0.0  & 0.75 & 0.0  & 0.65 \\ 
SJ\_D5.0      & Y & 1.0 & 5.0 & 0.0  & 0.75 & 0.0  & 0.68 \\ 
\cline{1-8}\\
\multicolumn{8}{c}{$\Delta$PA}\\
\cline{1-8}\\
SJ\_PA5       & Y & 1.0 & 3.0 & 0.0  & 0.75 & 5.0  & 0.62 \\ 
SJ\_PA10      & Y & 1.0 & 3.0 & 0.0  & 0.75 & 10.0 & 0.60 \\ 
SJ\_PA15      & Y & 1.0 & 3.0 & 0.0  & 0.75 & 15.0 & 0.58 \\ 
SJ\_PA20      & Y & 1.0 & 3.0 & 0.0  & 0.75 & 20.0 & 0.51 \\ 
\cline{1-8}\\
\multicolumn{8}{c}{Triaxial Atmospheres}\\
\cline{1-8}\\
TX\_E75\_1    & Y & 1.0 & 3.0 & 0.375& 0.75 & 0.0  & 0.65 \\ 
TX\_E75\_2    & Y & 1.0 & 3.0 & 0.50 & 0.75 & 0.0  & 0.72    
\enddata
\tablecomments{\label{atmostable} Runs in which the standard atmosphere was
varied one parameter at a time, holding the standard jet ($v_{\text{jet}} = 
100 c_s \exp[-3t]$, $\alpha = \pi/35$~rad and $\beta = \pi/15$~rad for 
$r_{\text{inner}} = 0.1$) constant.  The {\tt STANDARD} run uses the standard
jet and atmosphere and hence is a data point in each category.  The ``wings''
column denotes whether a run produced noticeable wings at any point during
its active lifetime.  $W/L$ at $t = 0.5$, on the other hand, is a way to
directly compare different runs.  At $t = 0.5$, $v_{\text{jet}} \sim 20c_s$,
i.e. twice the lobe material sound speed and a transition point during the 
active phase between a powerful and weak jet.  The time of this transition
depends on the velocity profile of the jet; $t = 0.5$ is only correct for the
standard jet.  $\Delta \text{PA}$ is the angular distance between the jet and
the major axis.}

\end{deluxetable}

\paragraph{{\it Fiducial Run}}
Our fiducial run ({\tt STANDARD}, Tables~\ref{atmostable} and \ref{jettable}) 
is the combination of 
our fiducial atmosphere ($r_0 = 1.0$, $\rho_0 = 3.0$, $\epsilon_x = 
\epsilon_y = 0.0$, $\epsilon_z = 0.75$) with our fiducial jet 
($v_{\text{jet}} = 100 c_s\exp[-3t] $, $\theta_{\text{jet}} = \pi/15$, 
$\alpha = \pi/35$, aligned along the $z$-axis).  Like most of our runs, this
simulation proceeds to a code time of $1.0$, at which point the radio galaxy
is in the early passive phase.  Unsurprisingly, this run produces some of the
most prominent wings of our suite (Figure~\ref{diffwings}); the shallow pressure gradient along the
$z$-axis and declining velocity profile combine to  
produce stalled lobes which begin to collapse at the end of the active phase.  
At the same time, wings quickly escape the core region, and by 
$t = 0.4$ are easily recognizable (right-hand panels of Figure~\ref{coevol}).  
After reaching $W/L = 0.83$ at $t = 0.75$, the cocoon collapse leaves the 
wings as the most notable features (Figure~\ref{standardgrowth}).  

At $t = 1.0$, we re-inject a jet with $v_{\text{jet}} = 100 c_s\exp[3(t-1.0)]$
(Figure~\ref{standardgrowth} and bottom panel of Figure~\ref{coevol}).  Since
the jet is expanding into tenuous material from the old cocoon, the expansion
quickly re-establishes a cocoon.  However, the reinjection also drives
a bow shock inside the collapsing cocoon, allowing initially strong back-flows to 
feed the wings directly.  Hence, between $t = 1.10$ and $t = 1.35$, an X-shaped 
radio galaxy is apparent ($W/L \sim 1.2$).  After $t = 1.35$, the jet again 
weakens to the point where collapse begins.  If the jet is instead reinjected
at the inception of the passive phase at $t \sim 0.75$, the end result is
very similar ($W/L \sim 1.2$ at $t = 0.95$).  

The atmosphere in this run is similar to the 2D simulations presented in \citet{capetti02} and
3D simulations in \citet{zanni03}.  Striking asymmetries develop despite the axisymmetric
atmosphere because of turbulent mixing between different slices in $\phi$ and
because the back-flows form three-dimensional structures within
the lobes.  If these flows are slightly misaligned on opposite sides of the
midplane, asymmetries develop.

\paragraph{\textit{Atmosphere}}
In our exploration of different atmosphere properties (Figure~\ref{atmosplots};
Table~\ref{atmostable}), we vary $\epsilon$, $\rho_0$, $r_0$, and
$\Delta \text{PA}$ individually while keeping the other atmosphere parameters
steady and using the standard jet ($v_{\text{jet}} = 100c_s\exp[-3t]$, 
$\rho_{\text{jet}} = 0.01$, $\alpha = \pi/35$, and $\beta = \pi/15$).  
Except when we vary $\Delta \text{PA}$, the jet is coaligned with the major 
axis.  

It is clear from Figure~\ref{atmosplots} that wing prominence depends strongly on 
the core radius $r_0$, core density $\rho_0$, and ellipticity $\epsilon$.  
Because the lobes and wings evolve almost independently, these 
dependencies can be understood by breaking the axial ratio $W/L$ into its
constituent parts: the wing length $W$ and lobe length $L$.  For the
purposes of this discussion, $W$ and $L$ are both taken to be measured at
$t = 0.5$ (Table~\ref{atmostable}). 

We find that $L$ varies only slightly as a function of $r_0$ (less than 10\%) but does
peak between $r_0 = 0.5$ and $r_0 = 2.0$.  At very low $r_0$, the lobes
are not confined by the atmosphere and spread out laterally, causing more
turbulent mixing and dissipating the thrust of the jet.  At very high $r_0$,
the lobes are completely confined by the atmosphere and the density gradient
is so shallow that as the jet weakens it is increasingly resisted by the
atmosphere.  The importance of this effect on the lobes is more pronounced in
a comparison between $L$ and $\rho_0$ where the lobe length declines 
by 50\% between $\rho_0 = 1.5$ and $\rho_0 = 5.0$.  

On the other hand, the wing length $W$ depends much more strongly on the
core radius $r_0$ (declining precipitously with increasing $r_0$) and only
weakly on the core density.  This behavior occurs because the wings respond
primarily to the pressure gradient along the minor axis (Equation~12) which
depends on $r_0$ and $\rho_0$ as $\nabla P \propto \rho_0 r_0^{-2}$.  
Thus, a large $W/L$ requires small, dense atmospheres.  

In this context, the importance of ellipticity is clear: $\epsilon$ determines
the ratio of core radii along the major and minor axes.  For high $\epsilon$,
the lobes expand into material which
does not vary much in density during the active phase whereas the wings quickly
escape the core region.  This behavior is unsurprising; \citet{capetti02}
and \citet{kraft05} make the same basic argument for wings produced when the
jet is coaligned with the major axis of the atmosphere.  However, 
the interplay between the 
parameters rules out predicting the wing length from first principles.  For
instance, one might imagine that we could produce wings in an atmosphere with
$\epsilon \sim 0.3$ (Figure~\ref{atmosplots}) by making $r_0$ tiny and 
$\rho_0$ large.  Although this \textit{does} increase the wing length, 
the radio galaxy expands beyond the dense part of the atmosphere 
during the ignition phase so the jet does not face as much resistance
as might be expected. 

It is possible to produce wings in atmospheres with smaller $\epsilon$ by making the
atmospheres triaxial (using prolate ellipsoids) instead of axisymmetric.  
\citet{zanni03} expect triaxial atmospheres 
\citep[as opposed to the axisymmetric simulations of][]{capetti02} to produce
longer wings because the backflow is collimated along a single minor axis
as opposed to forming a torus at the midplane.  We find that while a triaxial atmosphere
does collimate the wings, it does not by itself make them substantially longer
relative to an axisymmetric atmosphere with the same maximum $\epsilon$ 
because the backflow in our simulations is a compressible fluid and wing 
growth is driven by buoyancy (the pressure gradient is the same along the
minor axis).  Rather, 
triaxial atmospheres produce better-defined proto-wing channels during the ignition
phase.  These channels are then reinforced during the active phase.  In
other words, triaxiality increases $W/L_{0.5}$ at smaller values of $\epsilon$
because the ignition cocoon is less axisymmetric.

Finally, we test the sensitivity of wing formation to the degree of alignment 
between the jet and major axis $\Delta \text{PA}$ (Figure~\ref{atmosplots}d).  
Long wings are difficult to produce when the jet is 
misaligned with the major axis, and the wings produced differ in 
character.  This is because the ignition cocoon produces proto-wings
perpendicular to the jet instead of along the atmosphere's minor axis.  
Therefore, the longer wings which develop during the active phase do not 
benefit as much from the initial supersonic expansion.  Wings also become
increasingly associated with the 
lobe on their side of the major axis as $\Delta \text{PA}$ increases because
the lobes are bent by the atmosphere and the back-flows no longer make it
all the way to the midplane before flowing into the wings.  Hence, the 
lobe-wing pairs are mirror-symmetric about the midplane.  The angle between
the lobe and the wing in each pair is also largely determined by 
$\Delta \text{PA}$.

\setlength{\tabcolsep}{0.03in} 
\begin{deluxetable}{lcclccc}
\tablenum{2} 
\tabletypesize{\small}
\tablecaption{Varying the Jet in the Standard Atmosphere} 
\tablewidth{0pt} 
\tablehead{ 
\colhead{Name} & \colhead{Wings?} & \colhead{$\rho_{\text{jet}}$} & 
\colhead{$v_{\text{jet}}(t)/c_s$} & \colhead{$\alpha$} & \colhead{$\beta$} &
\colhead{$W/L$} \\  &  &  &  &  &  & ($t=0.5$) 
}
\startdata
\multicolumn{7}{c}{Standard Jet}\\
\cline{1-7}\\
STANDARD      & Y & 0.01 & $100\exp[-3t]$   & $\pi/35$ & $\pi/15$ & 0.62 \\ 
\cline{1-7}\\
\multicolumn{7}{c}{Velocity Profiles}\\
\cline{1-7}\\
SA\_V30       & N & 0.01 & 30               & $\pi/35$ & $\pi/15$ & 0.39 \\ 
SA\_V50       & Y & 0.01 & 50               & $\pi/35$ & $\pi/15$ & 0.43 \\ 
SA\_V80       & Y & 0.01 & 80               & $\pi/35$ & $\pi/15$ & 0.44 \\ 
SA\_V100      & Y & 0.01 & 100              & $\pi/35$ & $\pi/15$ & 0.47 \\ 
SA\_VE1       & Y & 0.01 & $100\exp[-t]$    & $\pi/35$ & $\pi/15$ & 0.48 \\ 
SA\_VE6       & Y & 0.01 & $100\exp[-6t]$   & $\pi/35$ & $\pi/15$ & 0.67 \\ 
SA\_VG3       & Y & 0.01 & $100\exp[-3t^2]$ & $\pi/35$ & $\pi/15$ & 0.47 \\ 
SA\_VG6       & Y & 0.01 & $100\exp[-6t^2]$ & $\pi/35$ & $\pi/15$ & 0.51 \\ 
\cline{1-7}\\
\multicolumn{7}{c}{Jet Density}\\
\cline{1-7}\\
SA\_D0.005    & Y & 0.005& $100f\exp[-3t]$  & $\pi/35$ & $\pi/15$ & 0.73 \\ 
SA\_D0.05     & N & 0.05 & $100f\exp[-3t]$  & $\pi/35$ & $\pi/15$ & 0.19 \\ 
SA\_D0.10     & N & 0.10 & $100f\exp[-3t]$  & $\pi/35$ & $\pi/15$ & 0.09 \\ 
\cline{1-7}\\
\multicolumn{7}{c}{Jet Width}\\
\cline{1-7}\\
SA\_B30       & Y & 0.01 & $100f\exp[-3t]$  & $\pi/35$ & $\pi/30$ & 0.51 \\ 
SA\_B7.5      & Y & 0.01 & $100f\exp[-3t]$  & $\pi/35$ & $\pi/7.5$& 0.76 \\ 
SA\_B5.0      & Y & 0.01 & $100f\exp[-3t]$  & $\pi/35$ & $\pi/5.0$& 0.87    
\enddata
\tablecomments{\label{jettable}The standard atmosphere is a relaxed, isothermal
$\beta$-model with core radius $r_0 = 1.0$, core density $\rho_0 = 3.0$, 
ellipticity $\epsilon = 0.75$, and with the jet oriented along the major axis.
Few runs are needed to deduce the dependence of the morphology on jet parameters
other than the kinetic luminosity as a function of time, but in these cases
we vary $v_0$ by some factor $f$ such that $L_{\text{kin}}(t)$ 
is the same as in the standard atmosphere; see text for caveats of this approach.
}
\end{deluxetable}
\setlength{\tabcolsep}{0.05in} 

\paragraph{\textit{Jet}}
In contrast to the atmosphere parameter exploration, determining the dependence
of $W/L_{0.5}$ on the character of the jet is difficult because of the potential
for time dependence in the jet power.  This time dependence, along with the small
width of the jets, also makes it more difficult to test our findings 
observationally.  We insist that the jets be light, hypersonic flows 
(Section~2) which produce sources which resemble double-lobed radio galaxies,
and within these constraints test the dependence of $W/L_{0.5}$ on jet power as
a function of time, jet width, and density of jet material (Figure~\ref{jetplots}).
For each of the simulations in Figure~\ref{jetplots} we use the
standard atmosphere with $r_0 = 1.0$, $\rho_0 = 3.0$, $\epsilon = 0.75$, and
$\Delta \text{PA} = 0.0^{\circ}$.  Note that Figures~\ref{jetplots}a and
Figures~\ref{jetplots}b do \textit{not} conserve kinetic luminosity between
runs, whereas Figures~\ref{jetplots}c and Figures~\ref{jetplots}d do.  The
exploration of jet velocity is conducted with jets with the standard width of 
$\beta = \pi/15$ ($\sim 0.01$ code units wide) and density of 
$\rho_{\text{jet}} = 0.01$.

Figures~\ref{jetplots}a and \ref{jetplots}b demonstrate the importance of
decaying jets to long wings.  In Figure~\ref{jetplots}a, we use jets with
differing velocities but no time dependence, finding that none produce a
large $W/L_{0.5}$. 
$W/L_{0.5}$ actually \textit{increases} with increasing $v_{\text{jet}}$ even
though the faster jets also punch through the atmosphere more quickly.  This
is because the overpressured cocoon produced by a weaker jet during the 
ignition phase is less overpressured and thus produces smaller proto-wings.
Below $v_{\text{jet}} \sim 20c_s$, the sources are not cocoon-bounded at all.
Since the jets do not decay, $W/L$ obviously decreases with time.

Powerful jets which decay (Figure~\ref{jetplots}b) are effective at producing
wings because the proto-wings form during the ignition phase and the jet 
generates strong back-flows early on when the jet head is close to the 
midplane.  Since the lobes grow increasingly slowly, the subsonically 
expanding wings keep pace with the active lobes more
easily.  As is clear in Figure~\ref{jetplots}b, the slower the decay, the
smaller $W/L_{0.5}$ (diamonds represent exponentially decaying jets and triangles
jets with $v_{\text{jet}} = v_0 \exp[-at^2]$).  Jets with \textit{increasing}
velocity do not produce long wings because a powerful jet is necessary at the
inception of activity to produce proto-wings.

In Figures~\ref{jetplots}c and Figures~\ref{jetplots}d we investigate the
dependence of $W/L_{0.5}$ on jet width and density.  We vary $\beta$ and
$\rho_{\text{jet}}$ while conserving $L_{\text{kin}}$ by varying the initial
velocity of the jet $v_0$, using the standard form of 
$v_{\text{jet}} = v_0 \exp[-3t]$.  We limit these runs to those with $v_0$
capable of producing an overpressured cocoon that seeds proto-wings.

The nozzle width of the jet clearly has a strong effect on $W/L_{0.5}$ 
(Figure~\ref{jetplots}c). Increasing
the jet width makes the ignition cocoon larger and more overpressured during
the ignition phase, promoting expansion along the minor axis.  The lobe
width during the active phase increases with increasing $\beta$, dissipating
the jet thrust over a larger solid angle.  Short, fat lobes attached to fat
wings result.  On the other hand, if $\beta$ is tiny, the jet drills through
the surrounding atmosphere quickly, and thin, long lobes result.  It is not
clear whether jet widths actually vary substantially between sources and what
determines the width of the jet; all jets are very narrow.  Thus, our
results are more generally a statement that jets which dissipate their thrust
over a wider area produce longer wings.

Figure~\ref{jetplots}d shows $W/L_{0.5}$ as a function of $\rho_{\text{jet}}$.
The sharp decline with increasing $\rho_{\text{jet}}$ is due to the thrust
carried by the jet.  Even at moderate velocities, denser material 
drives the lobes forward much faster than light material while at the same
time driving weaker back-flows.  Hence, wing formation is not favored.  We
note that when $\rho_{\text{jet}} > 0.05$, the sound speed of the lobe
material is not much greater than that of the ambient medium and the jet 
does not develop the usual KH instabilities.  These jets are therefore not
``light'' as required to reproduce realistic radio sources with non-relativistic
hydrodynamics (Section~2).  However, these runs are shown along with the light
ones to illustrate the importance of dissipating jet thrust to long wings.

\begin{deluxetable*}{lccccclccc}[t]
\tablenum{3} 
\tabletypesize{\scriptsize} 
\tablecaption{Synthesis Runs} 
\tablewidth{0pt} 
\tablehead{ 
\colhead{Name} & \colhead{$r_0$} & \colhead{$\rho_0$} & \colhead{$\epsilon_x$} & 
\colhead{$\epsilon_z$} & \colhead{$\rho_{\text{jet}}$} & \colhead{$v_{\text{jet}}/c_s$} &
\colhead{$\alpha$} & \colhead{$\beta$} & \colhead{$W/L_{\text{max}}$}
}
\startdata
\multicolumn{10}{c}{Interpolation}\\
\cline{1-10}\\
WIDE\_VE1          & 1.5 & 4.0 & 0.0  & 0.75 & 0.01 & $100\exp[-t]$   & $\pi/90$ & $\pi/7.5$ & 0.58 \\ 
WIDE\_VE3          & 1.5 & 4.0 & 0.0  & 0.75 & 0.01 & $100\exp[-3t]$  & $\pi/90$ & $\pi/7.5$ & 0.73 \\ 
TX\_E75\_VE1       & 1.0 & 3.0 & 0.375& 0.75 & 0.01 & $100\exp[-t]$   & $\pi/90$ & $\pi/7.5$ & 0.63 \\ 
TX\_E75\_VE1\_BIG  & 1.5 & 4.0 & 0.375& 0.75 & 0.01 & $100\exp[-t]$   & $\pi/90$ & $\pi/7.5$ & 0.68 \\ 
(restart at $t=1.0$)&    &     &      &      &      &                 &          &           & 0.78 \\ %
TX\_E75\_VE1\_B7.5 & 0.75& 3.0 & 0.375& 0.75 & 0.01 & $100\exp[-t]$   & $\pi/90$ & $\pi/7.5$ & 1.02 \\ 
TX\_E75\_VE3\_B5.0 & 1.5 & 4.0 & 0.375& 0.75 & 0.01 & $100\exp[-3t]$  & $\pi/90$ & $\pi/5$   & 0.82 \\ 
TX\_E60            & 0.75& 3.0 & 0.30 & 0.60 & 0.01 & $100\exp[-t]$   & $\pi/90$ & $\pi/7.5$ & 0.44 \\ 
TX\_E60\_WIDE      & 0.75& 3.0 & 0.30 & 0.60 & 0.01 & $100\exp[-t]$   & $\pi/90$ & $\pi/5$   & 0.67 \\ 
(restart at $t=1.0$)&    &     &      &      &      &                 &          &           & 0.97  \\ %
TX\_E60\_D5.0      & 0.75& 5.0 & 0.30 & 0.60 & 0.01 & $100\exp[-t]$   & $\pi/90$ & $\pi/7.5$ & 0.71 \\ 
TX\_E45\_D5.0\_WIDE& 0.75& 5.0 & 0.225& 0.45 & 0.01 & $100\exp[-3t]$  & $\pi/90$ & $\pi/5$   & 0.58 \\ 
\cline{1-10}\\
\multicolumn{10}{c}{Other}\\
\cline{1-10}\\
DISK\_VE1          & 2.0 & 3.0 & 0.25 & 0.50 & 0.01 & $100\exp[-t]$   & $\pi/90$ & $\pi/7.5$ & 0.60 \\ 
                   & 0.25& 8.0 & 0.0  & 0.90 &      &                 &          &           &      \\ %
DISK\_VE3          & 2.0 & 3.0 & 0.25 & 0.50 & 0.01 & $100\exp[-3t]$  & $\pi/90$ & $\pi/7.5$ & 0.83 \\ 
                   & 0.25& 8.0 & 0.0  & 0.90 &      &                 &          &           &      \\ %
TURBULENT\tablenotemark{a}          & 1.0 & 3.0 & 0.0  & 0.75 & 0.01 & $100\exp[-3t]$  & $\pi/90$ & $\pi/7.5$ & 0.64 \\ 
SHELL\_VE1\tablenotemark{b}         & 1.0 & 3.0 & 0.0  & 0.75 & 0.01 & $100\exp[-t]$   & $\pi/90$ & $\pi/7.5$ & 0.53 \\ 
SHELL\_VE3\tablenotemark{b}         & 1.0 & 3.0 & 0.0  & 0.75 & 0.01 & $100\exp[-3t]$  & $\pi/90$ & $\pi/7.5$ & 0.71 \\ 
INTERMITTENT       & 2.0 & 2.0 & 0.0  & 0.75 & 0.01 & $100\sin^2[50t]$& $\pi/90$ & $\pi/7.5$ & 0.44 \\ 
INTERMITTENT2      & 1.0 & 3.0 & 0.0  & 0.75 & 0.01 & $100\sin^2[6t]$ & $\pi/90$ & $\pi/7.5$ & 0.47 \\ 
SLOW\_START\_1\tablenotemark{c}     & 1.0 & 1.5 & 0.0  & 0.75 & 0.01 & $20$ \textit{to}         & $\pi/90$ & $\pi/7.5$ & 0.42 \\ 
                   &     &     &      &      &      & $100\exp[-t]$   &          &           &      \\ 
SLOW\_START\_2\tablenotemark{c}     & 1.0 & 1.5 & 0.0  & 0.75 & 0.01 & $20$ \textit{to}         & $\pi/90$ & $\pi/5$   & 0.45 \\ 
                   &     &     &      &      &      & $100\exp[-t]$   &          &           &      \\
SLOW\_START\_3\tablenotemark{c}     & 1.0 & 5.0 & 0.0  & 0.75 & 0.01 & $20$ \textit{to}         & $\pi/90$ & $\pi/5$   & 0.55 \\ 
                   &     &     &      &      &      & $100\exp[-t]$   &          &           &      
\enddata
\tablenotetext{a}{This is the standard run with axisymmetric Kolmogorov spectrum density perturbations introduced to the atmosphere.}
\tablenotetext{b}{Shells of material with a peak amplitude $\rho = 3.0$ were superimposed on the standard atmosphere; shells 
are generated along the $z$-axis by using 2D sine or sinc functions with a period of $z \sim 0.5$.}
\tablenotetext{c}{$v_{\text{jet}} = 20c_s$ from $t = 0.0$ to $t = 0.25$ and $v_{\text{jet}} = 100\exp[-(t-0.25)]$ thereafter.} 
\tablecomments{\label{synthesistable}
A representative sample of our non-systematic exploration of parameter space; 
not all runs attempted are included.  ``Interpolation'' runs refer to those
runs which are a natural extension of Tables~\ref{atmostable} and
\ref{jettable} whereas ``other'' runs include substantially different 
atmospheres and jet behaviors.  See text in Section~3.2 for discussion.
}
\end{deluxetable*}

\paragraph{\textit{Synthesis Runs}}
We have used the insights gained from examining the dependence of wing 
prominence on various atmosphere and jet parameters to new runs with 
complex atmospheres and jets in an attempt to produce long, realistic
wings.  These simulations (listed in Table~\ref{synthesistable}) are not
a systematic exploration of any phenomenon.  We offer a few brief observations
here. 

Triaxial atmospheres offer the best hope of making long wings at lower
$\epsilon$ (e.g. Figure~\ref{diffwings}), but still cannot produce substantial 
wings in our simulations for $\epsilon \lesssim 0.45$.  Several runs from a 
large suite of such simulations are listed in Table~\ref{synthesistable} with 
the prefix {\tt TX}.  Generally, they follow the same trends as described in
above; the importance of both favorable atmosphere and jet parameters to wings
is obvious in the simulation with $\epsilon_z = 0.45$ 
({\tt TX\_E45\_D5.0\_WIDE}), where a wide, decaying jet in a small, dense
atmosphere is required to produce wings comparable to that in the standard
atmosphere at lower ellipticity.  
Embedded disks of hot material (i.e. a thick disk; runs {\tt DISK\_VE1} and 
{\tt DISK\_VE3}), however, alleviate the problem, allowing prominent wings to 
grow in atmospheres with globally small $\epsilon$.  These disks are somewhat
denser than the larger ISM, so the ignition cocoon effectively encounters a 
small, dense, highly elliptical atmosphere during the ignition stage.  The
disks are blown apart by the blast wave from the ignition stage and might not
be observable once the radio galaxy has turned on.  

In most of our models, the ignition stage occurs in a smooth, relaxed medium.  In
real galaxies, the ignition stage would occur in the galactic center where
mergers, dynamical effects, and other phenomena associated with AGN can 
significantly disturb the ISM.  We do not attempt to model real galaxies, but
find that adding Kolmogorov spectrum turbulence to our atmospheres or small
bar-like perturbations to the underlying gravitational potential near the
nucleus does not have a large effect on the ignition stage.  As long as there
is a sufficient amount of ISM distributed around the nucleus, an overpressured
cocoon can form.  On the other hand, in runs where we begin with a weak
jet of $v_{\text{jet}} = 20c_s$ (twice the sound speed of the lobe material),
and no cocoon is formed, wings do not form (the {\tt SLOW\_START} runs in
Table~\ref{synthesistable}).  The backflow model has
difficulty producing wings when a weak AGN suddenly becomes more powerful
because an overpressured cocoon only forms when the jet is completely 
confined by the atmosphere.  These runs can, however, produce something
akin to Z-shaped morphology by virtue of the lobes escaping the densest
regions of the atmosphere before the powerful jet turns on.  

We have briefly investigated the potential for gas-rich ``stellar
shells'' from minor mergers located periodically along the major axis to
produce wings \citep[motivated by][]{gopal10a}.  We use a simple model in which
only the positive peaks of two-dimensional sine/sinc waves are added to the
density of an underlying elliptical atmosphere; the center is cut out.  The amplitudes
are set at a maximum of $\rho_0$.  
The shells do not significantly impact the formation of
wings emanating from the center of the galaxy, although they do resist the
jet, allowing for a higher $W/L$ (the {\tt SHELL} runs in 
Table~\ref{synthesistable}).  In the model of \citet{gopal10a}, wings
are instead produced near the site of the shell, and we cannot reproduce this.
Obviously, these runs are not a thorough exploration of the effect of stellar 
shells, especially if they actually bend jets (in which case they are beyond 
the scope of our models).  

Motivated by the importance of a reinjected jet (Section~3.1), we examine
the role of rapid intermittency (runs {\tt INTERMITTENT} in 
Table~\ref{synthesistable}).  By this we mean jets which experience 
multiple outbursts during the run and which are essentially in an ``on'' state 
or an ``off'' state.  Generally, we find that rapid intermittency has the 
effect of suppressing wing formation and results in much more regular cocoons 
(in terms of their projected morphology) than single outbursts.  On the other
hand, ragged cocoons are produced by long periods of dead time.  In neither
case are wings promoted, as each successive brief outburst deposits most of
its thrust at the ends of the lobes, far from the midplane.  The chief
reason why such intermittency does not produce prominent wings 
is that wings expand subsonically.  
Since the duty cycles of the intermittent jets are much shorter than 
the crossing time, the cocoon expands as if a moderately powerful jet of
constant velocity were powering it.  Unlike in single decaying outbursts,
the intermittent jets deposit most of their thrust far from the midplane.  On
the other hand, intermittency on the timescale of an $e$-folding time of a
decaying jet can be effective.

\section{Simulation Limitations}

Because we have only solved the equations of hydrodynamics in evolving our
simple models, it is worth considering the impact of additional complexity.  

\subsection{Missing Physics}
Our models do not include magnetic fields, special relativity, or radiative
losses, and our simple setup does not take into account complex jet or 
atmospheric structure, feedback, and other processes that may be
important to radio galaxy morphology.  In practice, we do not believe these
omissions invalidate our results.  For example, special relativity and 
magnetic fields must play crucial roles in determining the character and 
transverse structure of the jet, but we only insist that our simulated jets 
reproduce the collimation, hot spots, and back-flows of FR~II sources.  In 
other respects (e.g. radiative efficiency), the jet and AGN are inside a 
``black box.''  Because we are concerned with the behavior of lobe material,
we believe these omissions are justified.  

Likewise, radiative losses  
(depleting of lobe energy via synchrotron emission and inverse Compton 
scattering) only become important relative to adiabatic losses at late
times and are not in a position to influence the formation of wings.  Radiative
losses from the ICM are also irrelevant on the timescales of AGN outbursts,
so feedback (the connection of jet power to the amount of material crossing
the inner boundary) resulting from cluster cooling is unimportant on these
timescales.  As to feedback from \textit{backflow} \citep{antonuccio10}, there is no
obvious recipe to describe what happens to the mass flowing across the inner
boundary, since the inner boundary radius is much larger than the nuclear
engine, but as the back-flows straighten, less material will have velocity
vectors pointing towards the origin and eventually the AGN would
cease to be ``fed'' by backflow.  This is consistent with our entirely 
artificial recipe for decaying  jets.

On the other hand, magnetic fields (that must be injected with the jet) and
relativistic jets (which have a higher thrust for a given mass-energy
density) may strongly influence the behavior of lobe material.

Toroidal magnetic fields
may help collimate the backflow and retard its mixing \citep[e.g.][]{braithwaite10}.
The apparent continuity of fields in some wings \citep[e.g. NGC~326 in][]{murgia01}
is suggestive, especially considering that in our models the lack of
collimation leads to wide wings.  However, \citet{huarte-espinosa10}
find in simulations of FR~II sources that turbulence sets in within the cocoon
even when ordered fields existed earlier, so it is unclear that the
magnetic fields in the wings are actually toroidal.  As to the realism of the
\textit{jet}, it is not our goal to understand the jet physics in detail, but
we note the persistence of experimental hypersonic fluid jets in recent work
by \citet{belan11}, which the authors use to argue that magnetic collimation
is only required near the base of the jet.  

The importance of relativistic jets to the cocooon morphology is unclear.
\citet{kom96} find that, for jets matched by velocity, pressure, radius and
power, relativistic jets produce preferentially wider cocoons than 
classical ones.  The authors suggest that this effect may be accounted for by
the higher thrust in the relativistic case.  However, \citet{rosen99} find the 
opposite when using the same matching conditions: relativistic jets tend to 
produce narrower cocoons compared to classical jets (more specifically, jets 
with a higher Lorentz factor have smaller cocoons).  Without matched relativistic
simulations it is impossible to directly test the importance of special
relativity on our cocoon sizes, but it is remarkable that \citet{rosen99}
and \citet{kom96} agree that the qualitative nature of the cocoon is unchanged.
Including relativity may modify $W/L$ but would
not fundamentally change the appearance of our winged sources because 
relativistic physics in the lobes would not change the processes which produce
wings.  In other words, we believe that special relativity is a 
second-order effect akin to magnetic fields.

\subsection{Initial Conditions}

It is worth asking whether wings form only because our atmospheres are
``smooth'', i.e. because we have not included turbulence in the ICM or
structure near the center (where the environment is presumably complex).
This question is particularly important given
the observed interactions between jets and molecular clouds 
\citep[e.g.][]{ly05}, stellar shells \citep[e.g.][]{gopal10a}, gas in companion galaxies
\citep{evans08}, and more generally the complex nuclear environments of
active galaxies \citep[e.g.][]{rosario10}.  To this end, we have introduced
density perturbations mimicking stellar shells along the major axis at 
various intervals, tested the impact of a single ring near the nucleus,
and introduced Kolmogorov-spectrum turbulence to the ICM.  None of these
structures alter the same basic evolution of winged sources.
Additional jet physics would need to
be in place in order to determine whether the jet itself can be bent by 
interaction with high density pockets \citep{gopal10a}.  All of 
these tests use density perturbations with a maximum amplitude
of the core density $\rho_0$.  Finally, bulk flows in clusters are 
clearly important to radio galaxy structure \citep{morsony10}, although most
XRGs are large, strongly bridged sources in which this may not be a defining
effect.  

\subsection{Artifacts}
It is encouraging that our simulated radio galaxies reproduce the basic 
features of other simulations in the literature \citep[e.g.][]{reynolds02,
heinz06,vernaleo07,antonuccio10}, but it is worth considering the impact of
boundary conditions and the jet structure on the morphology of the
lobes---our setup is designed to produce lobes like those of radio galaxies
with physics and conditions inherently different from those encountered in
nature.  The mixing experienced by the lobes is 
also artificial (naturally set by the size of the grid zones, which vary
along $r$ and $\theta$), but mixing is too slow a process to suppress the
formation of wings.  

The inner boundary is the most significant artifact of our simulation 
because it represents no physical analog but rather allows us to hide the
AGN and the jet collimation mechanism.  In addition, backflow crossing the
boundary slows down and disappears.  This has two consequences: not all
the backflow can be harnessed, and flows may be effectively directed
around the boundary by eddies as slowing backflow crosses the boundary.  In
the first case, the amount of material is to small to 
influence the wings.  In the second, we find that the inner
boundary must be small compared to $r_0$ (keeping the
physical size of the jets fixed) in order to prevent the initial cocoon from
being unduly influenced by the sphere.  Inner boundary
spheres which are too large tend to promote material flowing around them, hence
promoting early wings, but suppress midplane mergers and hence the prominence
of later wings.  The initial interstices also survive for longer for 
oversized $r_{\text{inner}}$.  For typical atmospheres 
($0.5 < r_0 < 1.0$), $r_{\text{inner}}$ must be somewhat smaller than 
0.1~code units; $r_{\text{inner}} = 0.05$ is a compromise between a small 
impact on the initial cocoon and reasonably sized grid zones.  Runs with
$r_{\text{inner}} = 0.01$ do not appear substantially different from 
those with $r_{\text{inner}} = 0.05$, and the
timestep is unreasonably small.  

We precess the jet injection footpoints rapidly
($20\pi$~Hz) in a very tight circle around the poles in order to break up the
jet symmetry and spread its thrust out over a larger working surface.  We only 
require that our jet reproduce internal features seen in other hydrodynamic
simulations along with the terminal shocks that give rise to the back-flows
and lobes.  Without breaking up the jets, the jet head travels very quickly and
produces unrealistic lobes \citep[e.g.][]{vernaleo06}; 
in MHD simulations we would expect helical 
instabilities to break up the jet.  Our precession scheme is thus an artifice 
that is motivated by observables but whose recipe is not an attempt to model 
a physical process \citep[c.f.][]{heinz06}.  However, one might worry about its
influence on backflow and wings because the precession angle is a free 
parameter (subject to the condition that $\alpha < \theta_{\text{jet}}$).  In
other words, we can tune the width of the lobes and the rate of growth of the
radio galaxy within some narrow range of parameters.  Since stalling the
jet head contributes to wing prominence via shortening the active lobes,
is the axial ratio $W/L$ artificially high?  We do not believe so.  Even with
tiny precession angles, we form long wings in highly eccentric atmospheres
(the only atmospheres where long wings form).  Moreover, the jet head advance
speed also depends on the kinetic luminosity as a function of time.  

\section{Properties and Predictions of the Backflow Model}

We have successfully produced winged and X-shaped sources solely by redirecting
back-flow with static environmental pressure gradients, and are now in a 
position to compare our simulated sources to observed XRGs (Figure~\ref{gallery}) in order to 
critique proposed wing formation mechanisms involving back-flows.  In these
models, wings are produced as plasma flowing back from the jet heads
is deflected into a direction misaligned with the jets.  How this deflection
occurs is uncertain \citep{leahy84,worrall95,capetti02,kraft05,gopal10a}, 
and until now the backflow model has not been rigorously investigated with 
three-dimensional simulations.  

Here we outline a series of predictions (expectations) for properties which
are, at least in principle, observable if our simulations are an adequate
representation of wing formation.  These predictions are also summarized in
Table~\ref{predictions}.  We then assess the backflow model in light of these
predictions and briefly discuss the implications of our work for the wider
sample of distortions to the canonical double-lobed morphology of
bridged radio galaxies.

\subsection{Predictions}

\paragraph{\textit{X-shaped and winged sources are one family}}
The most natural consequence of the backflow model is that short and long
wings are produced in the same way.  Short wings should then be more common
than long ones since the long wings are harder to produce.  Although this is
an obvious point, we note it as a potential observational test because
sources with shorter wings and the candidate X-shaped sources 
\citep{cheung07} have not been tested for several of the trends seen in 
XRGs \citep[most importantly, the jet--major axis 
correlation][]{capetti02,saripalli09}.  
Our simulated sources predict that winged sources will fall in line, albeit 
with a greater spread in parameter space.

\begin{deluxetable*}{llll}
\tablenum{4} 
\tabletypesize{\scriptsize} 
\tablecaption{Model Predictions} 
\tablewidth{0pt} 
\tablehead{ 
\colhead{} & \colhead{Expectation} & \colhead{Observed?} & \colhead{Reference}
}
\startdata
1. & Short wings common; (intrinsically) long        & No in SS09 sample & SS09  \\
   & wings rare                                      & (yes including C07?)&   \\
2. & Projection tends to enhance $W/L$               & Undetermined      &   \\
3. & $W/L$ correlated with $\epsilon_{\text{ISM}}$   & Sample size too small      & HK10  \\
   & (weaker correlation with host $\epsilon$)       & No?               & C02,SS09  \\
4. & Wings require a jet aligned near                & Yes               & C02,SS09,HK10 \\
   & the major axis                                  &                   &    \\
5. & $W/L$ enhanced by higher pressure,              & Yes?              & L10 \\
   & small atmospheres                               & Yes?              & SS09 \\
6. & Most XRGs should be weak FR~IIs                 & Yes               & C09,L10 \\
   & (and hence strongly bridged sources)            & Yes               & LW84 \\
7. & Intermittency on scales of 3--10~Myr            & Undetermined      &    \\
   & produces longer wings                           &                   &    \\
8. & Wings are fainter than lobes                    & Yes               & LW84 \\
9. & Wings have flat spectral indices                & Mixed results     & LR07 \\
10.& Backflow follows existing channels              & Yes?              & SS09  \\
   & (collimation requires existing channels)        &                   &     \\
11.& Wings grow subsonically                         & No? (large wings) &     \\
   & (hence wing AGN outbursts are old)              & No?               & DT02  \\
12.& Flow speeds in wings should be                  & Undetermined      &    \\ 
   & transonic for lobe material                     &                   &    \\
\cline{1-4}
   & Backflow model piggybacks on other hydrodynamic & Yes?              & SS09  \\
   & models                                          &                   &   
\enddata
\tablecomments{\label{predictions}Predictions for XRG properties if our model is accurate; note
that these may not hold for the ``backflow model'' generally as phrased in prior
work.  Predictions correspond to arguments made in Section~4.
}
\tablerefs{C02: \citet{capetti02}; C07: \citet{cheung07}; C09: \citet{cheung09}; DT02: \citet{dennett02}; 
HK10: \citet{hodges10a}; L10: \citet{landt10}; LW84: \citet{leahy84}; 
LR07: \citet{lal07}; SS09: \citet{saripalli09}}
\end{deluxetable*}

\paragraph{\textit{Projection almost always enhances $W/L$}}
Wings in our models (even those produced by triaxial atmospheres) tend to be
wider than the lobes and hold their shape better in rotation and projection.
Hence, we predict that projection almost always enhances $W/L$.  In other
words, some observed sources with high $W/L_{\text{obs}}$ probably have an 
intrinsically smaller aspect ratio (e.g. Figure~\ref{gallery}).  
If the projection angle can be worked out for a sufficient number of XRGs and 
winged sources, we expect the lobes in a number of XRGs to be shortened via 
projection, i.e. the intrinsic distribution of $W/L$ is
shifted from the observed distribution \citep[an idea of the observed
distribution may be found in][]{saripalli09}.  If this
does not turn out to be the case, the backflow model would need a collimation
mechanism (see subsonic expansion below) to be consistent with observations.
Unfortunately, this prediction makes it difficult to quantitatively compare
the simulated population to the observed on because in most XRGs the
projection angle is unknown.  

\paragraph{\textit{Long wings require high ellipticity}}
Our simulations predict that intrinsically long wings can only be produced 
in very elliptical atmospheres (Figure~\ref{atmosplots}).
The lowest $\epsilon$ for which we can make convincing wings is 
$\epsilon \sim 0.45$, and very long wings require $\epsilon > 0.55$.  
These values are much higher than the average powerful radio galaxy host 
\citep[$\epsilon_{\text{peak}} \sim 0.2$;][]{smith89} and even most XRGs.
A high initial atmospheric ellipticity (or higher order asymmetry) is a clear
prediction of our models, but confirming this behavior in observed sources is
difficult.  

For instance, a few sources with very small host $\epsilon$ and
large $W/L$ (the best example is NGC~326) appear to have wings originating
\textit{outside} the ISM, so it is not evident that the ellipticity of the ISM
is always the relevant value for comparison.  Further, the
ellipticity of the ISM may differ substantially from that of the host
galaxy on small scales \citep{diehl07} even when in broad agreement on galactic
scales \citep{hodges10a}.  In this case, the best example in the literature is
3C~403 (Figure~\ref{3c403}, where $\epsilon_{\text{ISM}} = 0.059$ and 
$\epsilon_{\text{optical}} = 0.25$ \citep{kraft05}.  Indeed, we find that 
elliptical atmospheres of moderate $\epsilon$ with embedded disks are as
effective at producing wings as single-component atmospheres with extreme
$\epsilon$, but the disks are largely destroyed by the radio galaxy.  High
resolution imaging may be required to see any disk remnants.  The triaxiality
of elliptical galaxies also plays a role, since we may not measure a true
ellipticity.  Finally, because projection tends to enhance $W/L$, extreme
values of $\epsilon \sim 0.75$ (as we use in the standard atmosphere) are
not required to produce observed long wings.  All else being equal, we
still expect the intrinsic $W/L$ to be correlated with $\epsilon$, but the
caveats outlined above make it a difficult proposal to test presently.

\paragraph{\textit{Wings require a jet pointed nearly along the major axis
of an anisotropic environment ($\Delta \text{PA} \sim 0$)}}
As expected \citep[based on the studies in][]{capetti02,saripalli09,hodges10a},
wings require a strongly asymmetric environment in which the jet is stalled
by progression along the long axis (Figure~\ref{atmosplots}).  The wings grow in the favorable
pressure gradients along the minor axis or axes.  Our models expect a fairly
strict requirement for the jet to be within $\sim 15^{\circ}$ of the major 
axis for substantial wings to be produced via the overpressured cocoon
channel (the usual method for seeding wings in our models).  
The requirement is even more stringent in the event that a disk is present, 
indicating that disks are likely to be important only in a minority of cases. 

As $\Delta \text{PA}$ increases, the wings become shorter (Figure~\ref{atmosplots})
as well as increasingly associated with the lobe on the same side of the
major axis.  These lobe-wing pairs thus have mirror symmetry about the
midplane.  This is in agreement with radio maps of such sources.  Since wings
long enough to qualify the source as ``winged'' can be produced for a
$\sim 15^{\circ}$ range of major axis--jet separation and truly X-shaped sources
are only (intrinsically) produced for $\Delta \text{PA} < 5^{\circ}$, we
would expect $\sim 2/3$ of winged sources to have lobe-wing pairs, with 
acute angles between a lobe and its wing produced for smaller angles and
obtuse angles for larger ones.  Thus, we suggest that the existence of a large
number of sources with lobe-wing pairs rather than true X-shaped morphology is
consistent with the backflow model.  However, we would also expect that 
wing length would decrease with increasing $\Delta \text{PA}$; this has not
yet been measured.

\paragraph{\textit{Longer wings are produced in higher pressure, smaller
atmospheres}}
Steeper pressure gradients occur for smaller core radius and higher core
density/pressure (Equation~12).  Since the steepness of the pressure gradient
influences the rate at which wings grow, these small, dense atmospheres (e.g.
the ISM) are better at producing wings.  Moreover, wings can only form via
the overpressured cocoon channel if the cocoon can escape the central core
region and its high density before it comes into pressure equilibrium.  The
top panels of Figure~\ref{coevol} demonstrate this point (see also
Figure~\ref{atmosplots}).  Finally, a high core density also resists the jet 
advance along the major axis, allowing the wings to grow longer. 
Thus, our models expect that the hosts of XRGs (where the jet is presumably
pointed along the major axis) have higher pressure, on average, than those 
normal radio galaxies with jet geometry favorable to wings.  \citet{landt10}
find that the nuclear regions of XRGs have high temperatures ($T \sim 15000$~K)
indicating that these regions may be overpressured.  Although this does not
directly correspond to our models of relaxed, isothermal atmospheres, our
requirement for a small, high pressure environment is in agreement with their
work.

\paragraph{\textit{Most XRGs should be weak FR~IIs}}
Another key ingredient in our models is a decaying jet that begins as a 
powerful FR~II source and decays to luminosities more typical of FR~I sources.  
A powerful jet is required to produce the overpressured cocoon and drive 
wings early on, whereas once wings begin expanding subsonically, a weakening
jet (which advances increasingly slowly along the major axis of the atmosphere)
allows the wings to become quite prominent.  If the atmosphere
is much smaller than the radio galaxy (as in many cases), a decaying jet will
still grow more and more slowly, allowing subsonically growing wings to
keep up.  This finding naturally explains
the observations that XRG radio powers tend to lie near the FR~I/II ``break''
\citep{cheung09} while also possessing strong bridges associated with powerful
FR~II sources \citep{leahy84}.  \citet{landt10} argue that XRGs are the
archetypal transition population between FR~I and FR~II sources, with about
half the XRGs in their sample having weak emission lines from the AGN 
(weak-lined FR~IIs are otherwise uncommon).  This suggests that there is indeed
an evolutionary progression in the AGN and jet luminosity of XRGs.  Our
models predict a fast-rise exponential-decay profile in which the XRG spends
most of its lifetime as a weak FR~II source (Figure~\ref{jetplots}).
However, we note that \citet{best09} find in an SDSS study that FR~I and II
sources are not as obviously separated as in \citet{ledlow96} and that FR~II
morphology occurs for a variety of radio powers and host galaxy masses.

Intermittency is another important prediction of our models for observed
XRGs, but the action of intermittency in our simulations may represent a 
more complex underlying process.  Intermittency on timescales similar to the
$e$-folding time of a decaying jet (near the early passive phase) is effective in our simulations
because it allows the jet to bypass the overpressured cocoon stage and form
a bow shock within the radio lobe itself, thereby depositing a large amount of
backflow into the wings from a very powerful jet close to the wing bases.
Reinjection is effective in a relatively narrow window of time (3--10~Myr
depending on the size of the radio galaxy, e.g. Figure~\ref{standardgrowth}) between the active and passive
phases, but this may be plausible: the ripples in the Perseus cluster \citep{fabian06} and Abell~2052
\citep{blanton09} do imply an AGN duty cycle of about 10~Myr.  This is 
consistent with our simulated sources between 50--100~kpc across.  Thus, it is
conceivable that regular intermittency is a viable mechanism for enhancing
wings, although we note that if the intermittency is very rapid or very slow,
it is ineffective (Section~3.2).  On the other hand, intermittency could be replaced in our models by
any mechanism which drives back-flows nearer the base of the wings such as the
motion of denser ISM into the path of the jet.  We note that intermittency may
also be effective for wings attached to the primary lobes farther from the
nucleus; a reinjected jet quickly re-establishes the cocoon and drives 
strong back-flows along its length as it rapidly reaches the prior hot spot.
If the jet reinjected is similar to the outburst which initially formed the
wings, we would still expect XRGs mostly to be weak FR~II sources.  

\paragraph{\textit{Wings should be fainter than the primary lobes}}
The active lobes of XRGs are typically brighter than the wings.  This behavior
persists at lower (MHz) frequencies and is therefore not
obviously attributable to spectral ageing, although spectral ageing may be in
play for some XRGs \citep{lal07}.  Our models expect the wings to be fainter 
because they are substantially wider than the collimated primary lobes yet
contain (in most cases) less lobe material.  We therefore suppose that the
wing material spreads out and decreases the magnetic field energy density
$U_B$.  

Assuming that $U_B$ decreases with increasing volume (i.e. assuming
something like equipartition conditions, although precise equipartition is
not required), the synchrotron emissivity in the wings will be substantially
smaller than that in the lobes.  In other words, ``spherical'' wings a factor 
of $\sim 2$ wider than the active lobes will have a factor of $\sim 8$ smaller
$U_B$ than in the lobes.  In 
optically thin conditions (a good assumption), the surface brightness obtained
from integrating through the wings only recovers a factor of $\sim 2$ in the
wings, so we expect them to be a factor of $\sim 4$ dimmer than the active
lobes (more generally, the square of the ratio between the wing width and the
lobe width) for electrons of the same $\gamma$.  Of course, in projection the
wings may appear even dimmer since the lobes will tend to be shortened and
therefore increase in surface brightness (ignoring relativistic dimming of
the counterjet) whereas the wings will not.  

Clearly, this prediction rests on a number of assumptions not included
in our models and is therefore somewhat weak.  Since the wings are filled
with turbulent plasma, it is also possible that the magnetic field strength
is increased by winding up of fields as the material mixes.  Moreover, the
jet power (and therefore the back-flow speed) is time dependent in our
simulations, so the ratio between the surface brightness of the lobes and
the wings would be time dependent in our model as well.  Nonetheless, the
diffusion of material in the wings and the aforementioned projection effects
seem likely to dim the wings in the backflow model.

\begin{figure*}[t]
\begin{center}
\includegraphics[scale=0.65]{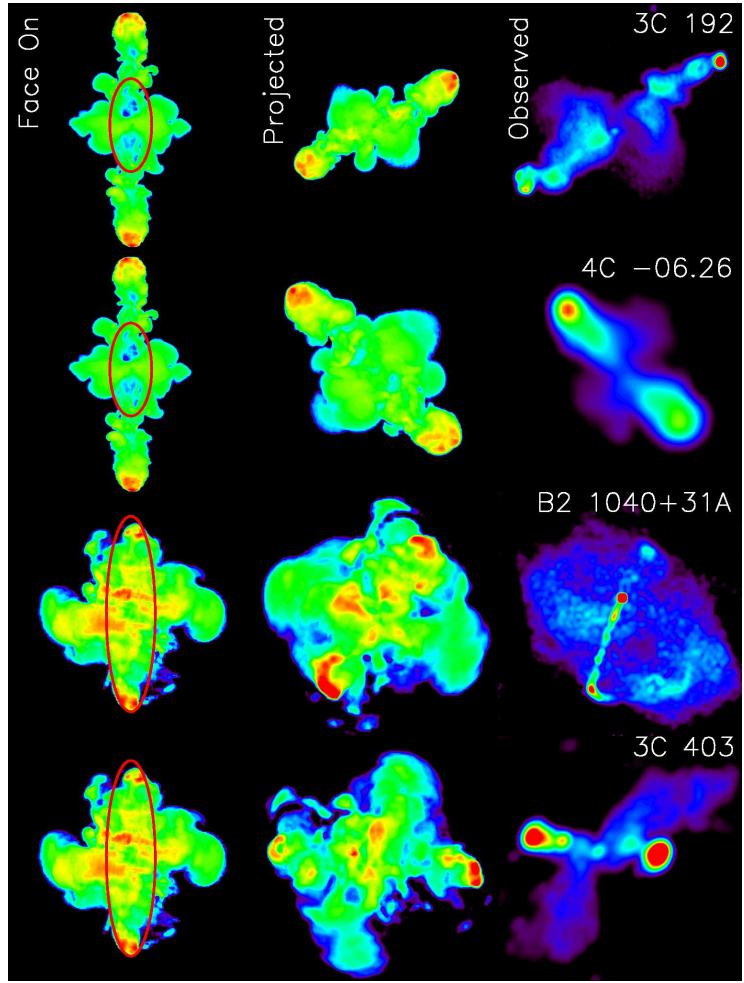}
\caption[]{{\small Gallery of false equipartition synchrotron maps 
($j_{\nu} \propto p/\rho^{7/4}$) showing the resemblance between our
winged sources and observed XRGs, and the importance of projection.  
Left: Face-on view showing intrinsic structure.  The red ellipses show the
core radius isobar of the model atmosphere.  Center: Rotated and
projected view.  Right: XRG analog (various resolutions). 
We only show a few examples of simulations and note that any one simulation 
can reproduce several observed sources; more than 60 of the sources in 
\citet{cheung07} can be plausibly reproduced by our simulations (in terms of 
appearance).}}
\end{center}
\end{figure*}

\paragraph{\textit{Backflow follows existing channels}}
Most of the wings in our simulated sources are seeded by the pressure-driven
expansion of an overpressured cocoon early in the source's life.  This cocoon
expands asymmetrically because of the asymmetric pressure gradients and forms
proto-wings which are later bolstered by the merger of laminar back-flows
near the midplane sending material into the wings during the active phase.  The
laminar back-flows themselves are ineffective at drilling new channels.
During the overpressured phase they acquire a vorticity near the
terminal shocks that leads them to follow the contact discontinuity of the
cocoon and ultimately flow back towards the AGN (Figure~\ref{backflow}).  During the active phase,
they are mostly straight and flow towards the midplane.  

Lobe material is also ineffective at producing channels because it follows
the path of least resistance: in a confined ellipsoidal cocoon, it will
simply spread out to increase the pressure throughout the cocoon rather than 
break out in a particular direction.  This can be seen in the false 
synchrotron maps in Figure~\ref{coevol}.  These maps assume equipartition 
and are not indicative of what the sources would look like in the GHz bands.
Rather, they effectively trace pressure in the cocoon, and it is easy to
see that the pressure in the confined cocoon on the left side of the
figure remains high relative to the X-shaped source (the false radio maps
use the same scale).  

Apart from proto-wings produced by the early pressure-driven expansions, 
Kelvin-Helmholtz and Rayleigh-Taylor instabilities produce channels which
backflow reinforces.  Given sufficient time and a powerful jet, these 
whorls and fingers can become wings in their own right, but likewise this is
due to the growth of the instabilities and buoyancy rather than 
redirected laminar back-flows.  

Notably, some sources exhibit wings which do \textit{not} emanate from the
center or are unlikely to have been produced via an overpressured cocoon
(e.g. NGC~326).  Because it is difficult to channel the back-flows, we
hypothesize that if the backflow model is correct, some seed proto-wing was
necessary to produce such sources.  The origin of these proto-wings is unclear.

\begin{figure*}[t]
\begin{center}
\includegraphics[scale=0.65]{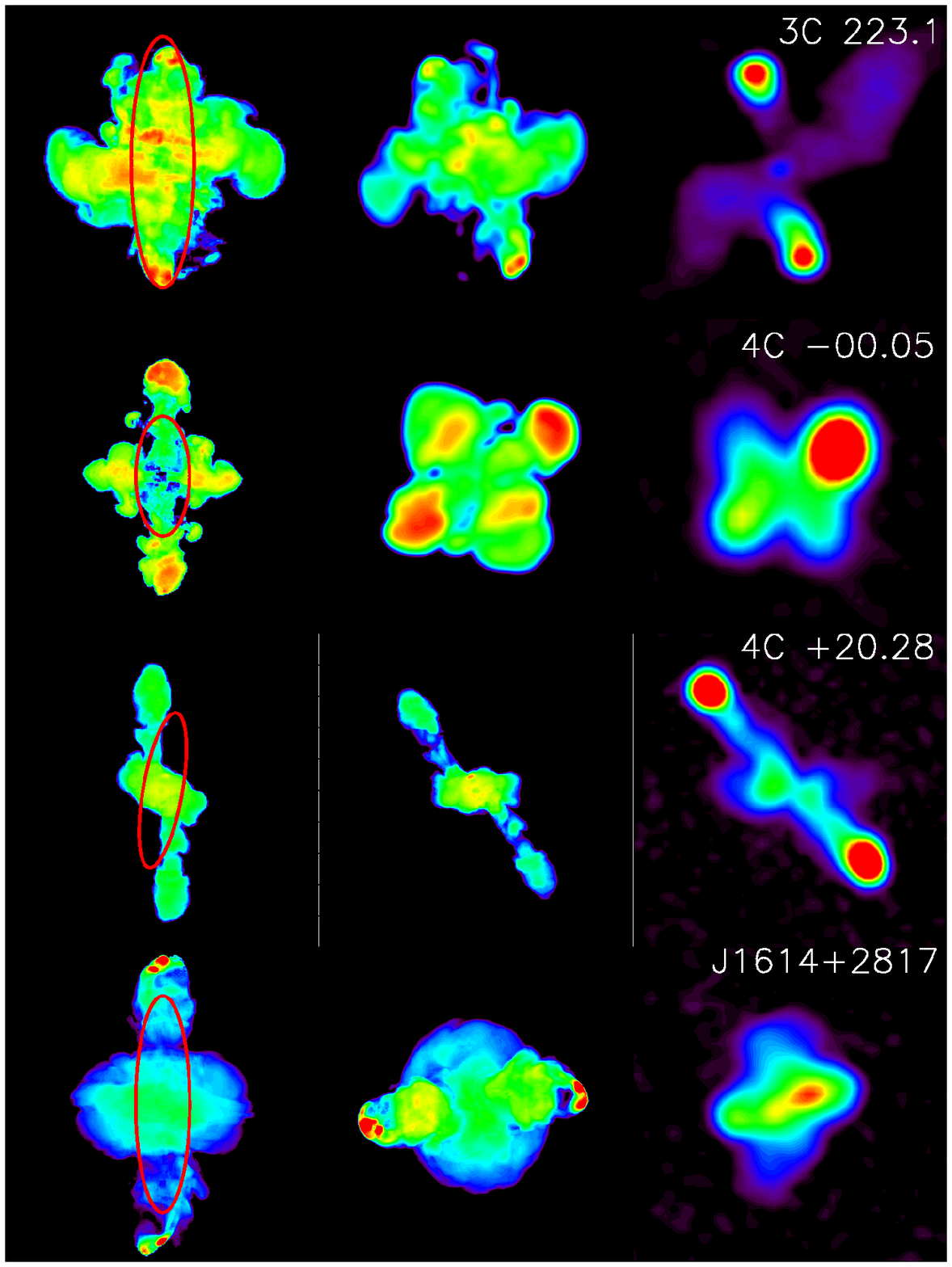}
\caption[]{{\small Figure~9 continued.}}
\end{center}
\end{figure*}

\paragraph{\textit{Wings expand subsonically}} 
Related to the expectation that back-flows are not directly responsible for
drilling channels is the prediction that the wings of radio galaxies
expand subsonically for most of their lifetimes (Figures~\ref{backflow} and
\ref{wingvel}).  This expectation is contrary to the \citet{capetti02} 
``overpressured cocoon'' and \citet{worrall95} or \citet{leahy84} 
``buoyant backflow'' proposals in the literature (Section~1) and poses a 
serious problem
for the backflow model in terms of explaining long wings.  We therefore
discuss this point in some detail below.

The problem of subsonic expansion can be broken into two distinct objections: 
(1) If the wings grow subsonically while the primary
lobes grow supersonically, how can wings be longer than the lobes? and (2)
Subsonic expansion implies wing lifetimes of more than 100~Myr for sources
hundreds of kpc across.  The first objection is easier to reconcile with
our models because our jets provide a natural mechanism for the subsonic
expansion of the lobes, and it is not clear whether most radio galaxies (even 
powerful ones) really do expand very supersonically.  Further, as we have noted, 
projection tends to enhance wings.  However, the lifetimes of large sources 
are more difficult to explain.

To see this, consider that for a rich cluster atmosphere
with $kT \sim 10$~keV (this corresponds to a $c_s \sim 1000$~km~s$^{-1}$ or,
conveniently, $1.0$~kpc~Myr$^{-1}$), subsonic growth implies that 
wings 100~kpc from base-to-tip would be at least
100~Myr old.  In group environments, where $c_s \lesssim 500$~km~s$^{-1}$
and monotonically declines at large radii \citep{sun09},
the longest observed wings (e.g. 3C~315 whose wings span $\sim 400$~kpc) 
could be almost 500~Myr old!  Even if AGN activity persisted this long, 
synchrotron cooling might set in: if the bulk flows of replenishing
backflow travel at a lobe sound speed $10c_s$, they would take between
20--100~Myr to replenish the leading edges of the wings.  Depending on the
break frequency and assuming magnetic fields in equipartition ($B$ of a few
$\mu$G) and a Lorentz factor $\gamma$ of several$\times 10^3$, this may
easily exceed the radiative cooling time.  

To solve this problem, variations of the backflow model require that the wings
actually do expand supersonically.  In the ``overpressured
cocoon'' model \citep{capetti02}, supersonic wing growth is achieved through
sustained pressure-driven expansion, whereas in the ``buoyant backflow'' 
model \citep{worrall95}, wings are driven by the nearly free expansion of 
diverted hypersonic back-flows.  Our models reproduce overpressured cocoons
and hypersonic back-flows, but the overall wing advance speed is subsonic.

For the overpressured cocoon model to produce long wings, a sustained 
overpressured state must be maintained to drive outflows.  However, in our
models the highly overpressured cocoon that forms soon after ignition
quickly expands to reach pressure equilibrium: the supersonic expansion
phase lasts at most around 15\% of the lifetime of the radio galaxy regardless
of the jet we inject (e.g. Figure~\ref{wingvel}).  As the radius of the (isobaric) cocoon grows, its
pressure falls much more rapidly than does that of the surrounding atmosphere
at a similar radius \citep[see also][Figure~2 in
their paper]{zanni03}, so the advance speed of the lateral cocoon expansion
falls precipitously from a peak of about Mach~2.  Thus, overpressured expansion
appears to be relevant only very early in the life of the source and by itself
can produce only short wings.  Indeed, this conclusion is supported by earlier
simulations \citep{zanni03} where even in advantageous triaxial atmospheres,
the overpressured cocoon produced at most an intrinsic $W/L$ of $\sim 0.5$
(the jets in their work were held at constant velocity and $W/L$ would thus
decrease over time).  \citet{saripalli09} suggest that the overpressured
state may instead result from backflow piling up upon reaching the central
(dense) region of the galaxy.  Our models do not support this scenario, since
the cocoon is contiguous and nearly isobaric throughout the active phase.

The buoyant backflow model argues that a combination of buoyancy forces and
wings driven by hypersonic flows produce large X-shaped sources.  For
instance, \citet{dennett02} estimate that the current outburst in 3C~403 
(Figure~\ref{3c403}) started
16~Myr ago.  If this outburst was solely responsible for generating the
$\sim 100$~kpc wings hydrodynamically \citep[as preferred by][]{kraft05},
the average expansion speed of the wings would have to be $\sim 8000$~km~s$^{-1}$
($0.027c$).  In contrast, a typical sound speed for galaxy groups is
$c_s \lesssim 500$~km~s$^{-1}$, a factor of $16$ smaller.  Likewise, 
\citet{worrall95} prefer hypersonic expansion.  What could drive such wings?
The answer invoked by these authors is redirected back-flows, which can be
accelerated at the terminal hot spots up to a few percent of $c$
\citep[based on the spectral ageing--distance method of][]{alexander87}.  The
observational inference of high backflow speeds is consistent with our models
and others \citep[e.g.][]{antonuccio10}, which find back-flow speeds of about
twice the lobe sound speed ($c_{s,\text{lobe}} = 10c_s$).  Further,
\citet{saripalli09} argue that the collimated morphologies of some wings 
requires some driving force.

\begin{figure*}[t]
\begin{center}
\includegraphics[scale=0.65]{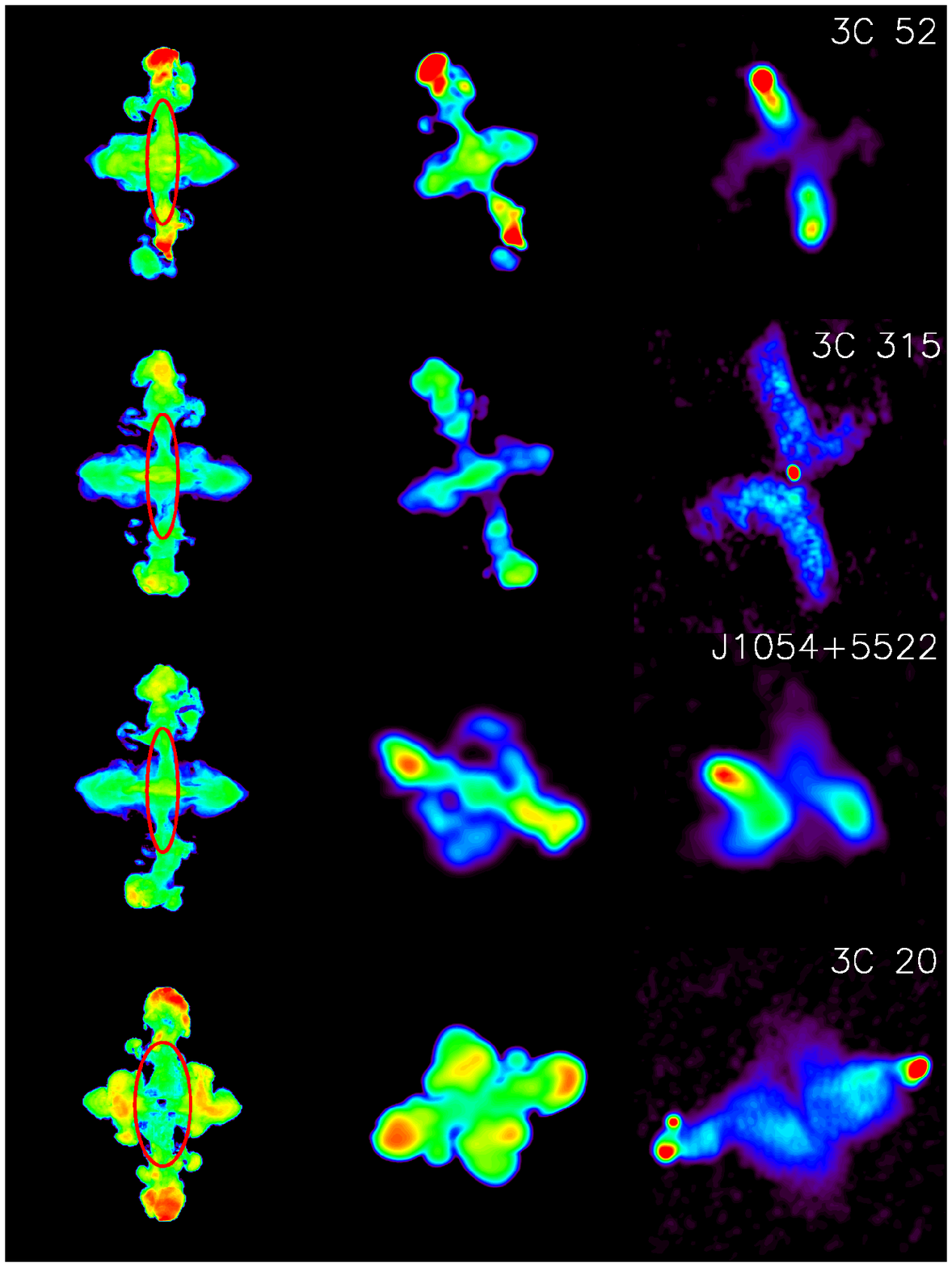}
\caption[]{{\small Figure~9 continued.}}
\label{gallery}
\end{center}
\end{figure*}

However, it is not evident that these laminar back-flows can be harnessed to drive
wing expansion (see the prior prediction).  
Subsonic wing growth in our simulations is a natural 
consequence of the tendency of backflow (lobe material) to mix and expand.
The fast back-flows from the terminal shocks merge near the midplane, 
dissipating their speed and driving flows into the wings which are subsonic
relative to the low-density material in the lobes (Figure~\ref{backflow}),
although they are still supersonic relative to the background. 
These flows then expand and decelerate further in the wings, dissipating their
thrust over a very large area.  Hence, detection of $10^3 - 10^4$~km~s$^{-1}$
flows in the wings would be insufficient evidence for hypersonic wing 
expansion.  Supersonic wing expansion would require collimation to prevent
thrust dissipation.  

Can a collimation mechanism be found?  An obvious possibility we have not included is
ordered magnetic fields, since particles can move along them much more easily
than across, but \citep{huarte-espinosa10} find that in FR~II sources, lobe
magnetic fields become turbulent on timescales of $\sim 10$~Myr.  
More generally, we note that (weak) shocks found around double-lobed radio sources
often imply cocoon expansion near Mach~1.  If a powerful jet cannot drive
highly supersonic expansion, it seems unlikely that less powerful, more
disorganized backflow could do so.  

Lastly, it is important to emphasize that subsonic expansion is only a serious 
problem for very large wings or very young AGN with a large $W/L$ ratio.  Our
simulations can reproduce most winged and X-shaped sources with subsonic
expansion if the AGN outburst is between 10--100~Myr old instead of 1--10~Myr.
These simulations are also not the final word, as they only solve the
equations of hydrodynamics in a relaxed, non-dynamic atmosphere.  

\subsection{Assessment}

Our results, taken in context, strongly implicate a hydrodynamic origin for
X-shaped sources and a common origin for winged and X-shaped sources.
We can produce bona fide XRGs (i.e. with intrinsic $W/L > 0.8$) with a single outburst
from a jet with a plausible time dependent kinetic luminosity on timescales 
broadly consistent with observed sources.  The radio galaxies we produce
are consistent with prior simulations, and faithful reproductions of observed
radio galaxies can be generated by tuning the viewing angle 
(Figure~\ref{gallery}).  These sources are also consistent with most 
observed properties of the XRG population outlined above, and our simulations
reproduce elements of the overpressured cocoon and buoyant backflow models
while relying solely on the interaction of the radio lobe material with an
anisotropic environment.  Hence, the backflow model remains
a strong contender for the origin of XRGs---it would be surprising if some XRGs
were \textit{not} produced in this manner.  

However, our simulations also present new challenges to the proposals in
the literature: we find that long wings require unusually elliptical 
atmospheres and expand subsonically, making very long wings difficult to
explain.  Having examined the deficiencies of our simulations, we cannot
identify an obvious internal remedy; we note that pre-existing channels are
necessary (in the backflow model) for some sources in the literature and
may be generally important. 

The final prediction of our simulations is therefore that 
the XRG population is heterogenous and \textit{the backflow model works 
in tandem with other hydrodynamic models rather than solely on its own.}  
In short, this is because our models make it too difficult to
produce XRGs given their observed frequency.  

Consider that winged and 
X-shaped sources make up 5--10\% of double-lobed radio sources.  If, as in our
models, X-shaped sources are produced by a fortunate coincidence of jet
geometry, ISM ellipticity, and jet power, we might imagine that the fraction
of X-shaped sources is some function of each of these variables.  In the
simplest case, where each of these factors contribute independently to 
promoting wings, the fraction of XRGs might look like
\begin{equation}
f_{\text{XRG}} \propto f_{\text{PA}}^{\alpha} f_{\epsilon}^{\beta} f_{\text{kin.} L}^{\gamma},
\end{equation}
where $f_{\text{XRG}}$ is the fraction of double-lobed sources which are
X-shaped and the other $f$ values represent the fraction of sources for each
variable which meet the threshold criterion for wings.  The exponents are 
unknown measures of the relative importance of each variable.
Obviously, the fraction of XRGs must be smaller than the fraction of sources
which meet any one criterion.  

Now, consider that virtually all double-lobed radio galaxies 
emanate from elliptical hosts, and have apparently random jet--major axis 
orientations \citep[as opposed to the very weakest radio-emitting AGN whose 
jets seem to be pointed along the minor axis of their hosts;][]{browne10}.  
Our models suggest that a jet within $\sim 15^{\circ}$ of the major axis of
its host is required to produce wings.  In a uniform distribution of radio
jet position angles, we would thus expect that 
$f_{\text{PA}} \sim 15/90 \approx 0.17$.  Assuming all powerful jets have
a kinetic luminosity function conducive to forming XRGs, we still require about half of
all sources where the jet is co-aligned with the major axis to have high
ellipticity (assuming the exponents are all equal to unity).  Given that 
ellipticity is clearly a very sensitive parameter and that the peak
$\epsilon$ of powerful radio galaxy hosts is far below the wing threshold
\citep{smith89},
it is clear that our simulations (in this oversimplified formulation)
underpredict the observed frequency of winged sources.  In other words, 
wings are \textit{too difficult} to make in our models.  

To reconcile this result with the observations, either our models must be
fundamentally deficient with respect to the behavior of backflow or they
require alternate mechanisms to form proto-wing channels such as the 
jet--stellar shell interaction model of \citet{gopal10a} or the 
jet--merging ISM explanation for Z-shaped sources in \citet{zier05}.
As we have seen, backflowing material reinforces any proto-wings and can
turn them into full-fledged wings if the original mechanism fails to do so;
making the channels in the first place is what the backflow model cannot 
easily do.  

Thus, we propose that the backflow model has a commensal relationship with
other hydrodynamic mechanisms for forming wings by reinforcing and growing
any pre-existing channels accessible to a jet pointed along the major axis
of its host galaxy.  These channels would naturally grow most easily along
the steepest pressure gradient.  In cases where the jet power and atmosphere
size match appropriately, these channels would be produced by the expansion 
of an overpressured cocoon as in our simulations.  In other cases, channels
could be produced by the interaction of the jet with structure in the ISM
or by minor mergers.  In this scenario, we would still expect to see most of
the predictions of our simulations listed above for the case of pure 
backflow in a relaxed atmosphere.  

\subsection{Distortions to the Canonical Double-Lobed FR~II Radio Galaxy}

At this point, it is worth revisiting the buoyant backflow model as phrased
in \citet{leahy84}, where XRGs are unified with other distortions to the
canonical FR~II model (about two thirds of strongly bridged radio galaxies
show central distortions) by the deflection of backflow around a denser
medium.  These distortions may also be related to radio galaxies with
interrupted bridges by the ``superdisk'' model \citep{gopal09}, in which gas
displaced by a galaxy merger is believed responsible for docking the lobes in 
an asymmetric manner.  

Because the backflow model as expressed in our simulations depends only on
a fortuitous combination of atmosphere morphology and jet behavior to
produce winged sources, we can indeed reproduce the basic bridge distortions
of \citet{leahy84}.  However, we cannot reproduce long Z-shaped sources and
some asymmetric distortions which do not appear to be the result of bulk 
motion or cluster turbulence.  These include sources with a single wing on 
one side, sources with wings which themselves bend dramatically, sources with 
one FR~I lobe and one FR~II lobe \citep[HYMORs;][]{gopal00}.  Again, the main
hindrance seems to be the inability to channel backflow; jets which bend
may solve this problem.  Although there are elements of the behavior of
backflow which we do not presently understand, we come to the same conclusion
as with X-shaped sources: the backflow model can account for other bridged
distortions only in tandem with another hydrodynamic mechanism.  

\section{Summary and Conclusions}

We have conducted a series of three-dimensional hydrodynamic simulations of
light, hypersonic jets to study the viability of the backflow model for
the formation of wings in X-shaped radio galaxies.  The XRGs seem to be a
population unto themselves, with characteristic environmental geometry, 
radio power, black hole mass, etc., and any successful model must account for
these peculiarities.  Our main results follow.

\begin{enumerate}

\item The jets, back-flows, and lobes in our simulations are similar to those
in the recent literature, giving us confidence in the usefulness of our
simulations as probes of the backflow model.

\item Wings in our models form in two stages: the establishment of channels or
proto-wings and then buoyant (usually subsonic) expansion.  Specifically,
our models corroborate the overpressured cocoon model of \citet{capetti02}
early on, but as cocoons quickly come to pressure equilibrium with their
surroundings, most of the wing length represents subsonic expansion
(Figure~\ref{backflow}).

\item We have produced prominent wings by geometry and radio power alone,
proving that the backflow model can, in principle, make X-shaped sources
(Figure~\ref{coevol} and \ref{gallery}).  Both the atmosphere and jet kinetic luminosity
as a function of time are crucial to forming X-shaped sources.

\item Long wings are produced in a relatively small portion of 
parameter space, requiring galaxies with high ellipticity, decaying jets,
proper jet orientation, and appropriate atmosphere size (Figures~\ref{atmosplots}
and \ref{jetplots}). 

\item The main challenges to the backflow model are the requirement for
high ellipticity and subsonic wing growth.  Adding additional physics is not
obviously helpful.  The backflow model seems to require an additional mechanism to 
make proto-wings which the backflow reinforces; in our models, we make
these channels by the initial expansion of an overpressured cocoon in an
anisotropic environment, but this cannot explain every XRG.  We cannot
form new channels solely by deflecting back-flows.

\item If the backflow model can overcome the issues noted above, it is a very
strong candidate for explaining X-shaped and other disturbed radio galaxies.
Our models naturally reproduce many of the characteristics of the XRG
population (Table~\ref{predictions}).

\end{enumerate}

There are several natural extensions of this work which promise to be 
fruitful.  First, adding magnetic fields and investigating other potential
collimating mechanisms may rule out or boost the backflow model depending on
the wing expansion speeds attained.  Second, our models rely on the formation
of channels misaligned with the jets; these have been proposed in several
other contexts as well to explain radio galaxy morphology.  The origin of
these channels is not known, and identifying and testing candidates would
be important for wing formation models.  Third, more realistic jets could
determine whether the backflow model can support both long wings and active
FR~II lobes with hot spots; the bending of the jets in particular is an
important issue.  Finally, models which produce the wings hydrodynamically
but not explicitly by the deflection of backflow \citep[e.g. the recently proposed
stellar shell model][]{gopal10a} are worth exploring by making the atmospheres
more complex in tandem with more realistic jets.

\acknowledgments
The authors thank the anonymous referee for helpful and clarifying comments. 
We acknowledge support from \textit{Chandra} grant GO011138A.  

\singlespace



\end{document}